\documentstyle[12pt,epsfig]{article}

\newcommand{\be}{\begin{equation}}
\newcommand{\ee}{\end{equation}}
\def\aprle{\buildrel < \over {_{\sim}}}

\begin{document}  
\topmargin 0pt
\oddsidemargin=-0.4truecm
\evensidemargin=-0.4truecm
\renewcommand{\thefootnote}{\fnsymbol{footnote}}
\newpage
\setcounter{page}{0}
\begin{titlepage}   
\vspace*{-2.0cm}  
\begin{flushright}
FISIST/11-99/CFIF \\
hep-ph/9907399
\end{flushright}
\vspace*{0.5cm}
\begin{center}
{\Large \bf Resonance Spin Flavour Precession and Solar Neutrinos} 
\vspace{1.0cm}

{\large 
Jo\~{a}o Pulido
\footnote{E-mail: pulido@beta.ist.utl.pt} and 
E. Kh. Akhmedov
\footnote{On leave from National Research Centre Kurchatov Institute,
Moscow 123182, Russia. E-mail: akhmedov@gtae2.ist.utl.pt}}\\
\vspace{0.05cm}
{\em Centro de F\'\i sica das Interac\c c\~oes Fundamentais (CFIF)} \\
{\em Departamento de Fisica, Instituto Superior T\'ecnico }\\
{\em Av. Rovisco Pais, P-1049-001 Lisboa, Portugal}\\
\end{center}
\vglue 0.8truecm

\begin{abstract}
We examine the prospects for the resonance spin flavour precession as a solution to
the solar neutrino problem. We study seven different realistic solar magnetic field 
profiles and, by numerically integrating the evolution equations, perform a fit of
the event rates for the three types of solar neutrino experiments (Ga, Cl and
SuperKamiokande) and a fit of the energy spectrum of the recoil electrons in 
SuperKamiokande. A $\chi^2$ analysis shows that the quality of the rate fits is
excellent for two of the field profiles and good for all others with $\chi^2 /d.o.f.$ 
always well below unity. Regarding the fits for the energy spectrum, their quality is 
better than that for the small mixing angle MSW solution of the solar neutrino
problem, at the same level as that for the large mixing angle MSW solution but worse 
than that for the vacuum oscillations one. The experimental data on the spectrum are 
however largely uncertain especially in the high energy sector, so that it is too
early yet to draw any clear conclusions on the likeliest type of particle physics
solution to the solar neutrino problem.       
\end{abstract}
\end{titlepage}   
\renewcommand{\thefootnote}{\arabic{footnote}}
\section{Introduction}

More than thirty years after its first recognition \cite{Davis}, the solar neutrino
problem has become well established on the grounds of theoretical \cite{TCL}-\cite{BTC}
versus experimental developments \cite{Hom}-\cite{SuperK}. The neutrino oscillation 
solutions to the anomaly, both matter \cite{MSW} and vacuum oscillations \cite{VO},
have received a great deal of attention \cite{hata}-\cite{ber} and
became the most popular ones. 
On the other hand, the solution based on the resonance spin flavour precession (RSFP) 
mechanism, proposed in 1988 \cite{LMA}, involves the simultaneous flip of both 
neutrino chirality and flavour \cite{SV}. It has been much less thoroughly
investigated, in particular concerning its parameter range predictions \cite{Tom} and 
the quality of the fits involved. 
In fact, it is only recently that the first detailed data analysis based on the
minimum $\chi^2$ approach has been performed in the framework of this mechanism 
\cite{GN}. 
One of the reasons for this comparative lack of popularity may be the large order
of magnitude of the neutrino magnetic moment required for a significant conversion 
and our relatively limited knowledge of the solar magnetic field. It should however
be noted that models for a large neutrino magnetic moment not conflicting with the 
smallness of neutrino mass already exist \cite{BFZ}, and there are solar model 
indications favouring a field as large as $3\times10^5$ G around the bottom of the
convective zone \cite{Parker}. As shall be seen, the required neutrino magnetic moment 
is, on the other hand, not inconsistent with astrophysical limits \cite{magmo}, 
especially taking into account the inherent uncertainties of the analyses.

One of the main motivations for investigating the RSFP mechanism is
that neutrinos provide a unique probe for the interior solar magnetic field, if
they have a sizeable magnetic moment. In fact, previous analyses \cite{Tom},
\cite{ALP1}-\cite{ALP2} favour solar magnetic fields rising 
along a relatively short distance (a factor of at least 6-7 over a 7-9\% 
fraction of the solar radius at the most). This rise may be very sharp and even
discontinuous profiles are allowed by the data. This
appears as a natural consequence of the strong suppression for the intermediate energy
neutrinos ($^7$Be and CNO fluxes) together with the almost no suppression for the
low energy pp ones. Solar physics arguments suggest that such a sharp rise,   
if it exists, must lie around the upper layers of the radiative zone and
the bottom of the convective zone. Moreover the moderate suppression of the energetic 
$^8$B neutrinos favours a gradual decrease of the field along the convective zone,
with a larger slope at a greater depth.

In this paper we perform an investigation of the RSFP scenario
for seven different magnetic field profiles all obeying the general features
described above. We start with the numerical integration of the evolution equations
for two neutrino flavours, from which we obtain the survival probability as a function 
of energy. This is used to obtain the event rate in each of the three types of experiments
(Gallium, Chlorine and SuperKamiokande). The rates thus obtained are then confronted with   
the experimental ones \cite{Hom}-\cite{SuperK}. We let the parameters $\Delta m^2_{21}$
(mass square difference between neutrino flavours) and $B_0$ (value of the field at
the peak) run over their ``plausible'' ranges, namely $4\times10^{-9}$
eV$^2$$\aprle\Delta m^2_{21}\aprle 20\times 10^{-9}$ eV$^2$ and $3\times10^4G\aprle 
B_0\aprle 3\times10^5$ G. These ranges are ``plausible'' 
in the sense that they are dictated by consistency with the RSFP
scenario. The first is required by the location of neutrino resonances as a function of 
energy, so that the most strongly suppressed neutrinos have their resonance around the 
strongest magnetic field, and the second range by the solar physics requirement of keeping  
$B_0$ not greater than $3\times10^5$ G at the bottom of the convective zone but
large  
enough to produce significant conversion. For each magnetic field profile
we select those intervals in which the theoretical ratios approach their experimental
values, and perform a $\chi^2$ analysis in order to determine the local minima and the
best fit. We thus get impressively small values of $\chi^2/d.o.f.$, 
in the range $(3-5)\times
10^{-2}$ at the best fit in three of the profiles, two of which are remarkably stable
against variations of the parameters $\Delta m^2_{21}$ and $B_0$. We also compare the    
theoretical prediction for the recoil electron energy spectrum in SuperKamiokande 
in each best fit case with the experimental result.

The paper is organized as follows: in section 2 we present the magnetic field 
profiles to be investigated and the survival probabilities obtained from them upon 
integration of the evolution equation. In section 3 we calculate the ratios of the event
rates predicted in the RSFP scenario 
to the standard solar model (SSM) event rate in each experiment and analyse their dependence
on $B_0$ and $\Delta m^2_{21}$. Next we describe our $\chi^2$ analysis for the rates and  
identify the local and absolute minima in each field profile in terms of these parameters. 
The $\chi^2$ analysis for the electron recoil spectra in SuperKamiokande ends this section.
Finally in section 4 we present our summary and overview.

\section{Magnetic Field Profiles and Survival Probabilities}

In this section we will present the magnetic field profiles to be used throughout,
all satisfying the general features explained in the introduction, with a sudden rise
around the bottom of the convective zone, at approximately 0.71 of the solar radius, and
a smoother decrease up to the surface. Here the field intensity should not exceed a few
hundred Gauss. All profiles analysed are displayed in figs. 1 and 2. The first two
are simple triangle profiles, with the field intensity rising linearly from zero at 
fraction
$x=x_R$ of the solar radius, reaching a peak at $x=x_C$ and decreasing linearly to
zero at the solar surface \cite{ALP1} 
\be
B=0~~~,~~~x<x_R
\ee
\be
B=B_0\frac{x-x_R}{x_C-x_R}~,~x_{R}\leq x\leq x_{C}
\ee
\be
B=B_0\left[1-\frac{x-x_C}{1-x_C}\right]~ ,~x_{C}\leq x \leq 1
\ee
with units in Gauss. We study this class of profiles in two cases 
shown in fig. 1: profile 1, $x_R=0.70,~x_C=0.85$ (equilateral triangle) 
and profile 2,  
$x_R=0.65,~x_C=0.80$ \footnote{A slight variant of this second profile 
with a vanishing field for $x\geq 0.95$ was used in ref. \cite{GN}.}.

Next we consider a slightly larger upgoing slope and a quadratic like decrease up to 
the surface. This is profile 3 with $x_R=0.65,x_C=0.75$ (fig. 1)
\be
B=0~~~,~~~x<x_R
\ee
\be
B=B_0\frac{x-x_R}{x_C-x_R}~,~x_{R}\leq x\leq x_{C}
\ee
\be
B=B_0\left[1-\left(\frac{x-0.7}{0.3}\right)^2\right]~,~x_{C}<x\leq 1.
\ee
The two cases with an infinite slope at $x=x_R$ are considered next. The first is 
profile 4 with the same quadratic like decrease from $x\geq x_R$ up to the surface
\cite{ALP1} (fig. 1)
\be
B=0~~~,~~~x<x_{R}
\ee
\be
B=B_0\left[1-\left(\frac{x-0.7}{0.3}\right)^2\right]~,~x\geq x_{R},
\ee
with $x_R=0.71$ and the second is profile 5 with a reversed shape slope: a strong 
decrease at first, which then becomes quite moderate on approaching the surface 
\cite{AB,ALP2} (fig. 2) 
\be
B=0~~~,~~~x<x_{R}
\ee
\be
B=\frac{B_0}{\cosh30(x-x_R)}~,x\geq x_{R}
\ee
with $x_R=0.71$. Profile 6 is a modification of the previous one with a large
finite positive slope across the bottom of the convective zone \cite{Tom} (fig. 2)
\be
B=2.16\times10^3~~,~~x\leq 0.7105
\ee
\be
B=B_{1}\left[1-\left(\frac{x-0.75}{0.04}\right)^2\right]~,~0.7105<x<0.7483
\ee
\be
B=\frac{B_{0}}{\cosh30(x-0.7483)}~,~0.7483\leq x\leq 1
\ee
with $B_0=0.998B_1$. The last case considered is profile 7, similar to the 
previous one but with 
a linear decrease towards the surface \cite{Tom} (fig. 2)
\be
B=2.16\times10^3~~,~~x\leq 0.7105
\ee
\be
B=B_{0}\left[1-\left(\frac{x-0.75}{0.04}\right)^2\right]~,~0.7105<x<0.7483
\ee 
\be
B=1.1494B_{0}[1-3.4412(x-0.71)]~,~0.7483\leq x\leq 1.
\ee
The reasons to select these classes of solar field profiles are twofold: first to 
investigate previously proposed cases in the light of the 
most recent experimental data and second to explore those profiles which provide the
best and most stable fits against variations of $\Delta m^2_{21}$ and $B_0$. 
While, as will be 
seen, profile 3 provides the best fit of all, this is only marginally better
than the fits corresponding to profiles 5 and 6 which are on the other hand much more
stable, clearly favouring a specific class of field shape. 

We consider the neutrino propagation equation through solar matter which for the two
generation Majorana case and zero vacuum mixing angle reads \cite{PhysRep,Heid}
\be
i\frac{d}{dt}\left(\begin{array}{c}\nu_{e_L}\\ \bar\nu_{\mu_R}\end{array}\right)=
\left(\begin{array}{cc}V_e&\mu_{\nu}B\\
\mu_{\nu}B&\frac{{\Delta m^2}_{21}}{2E}+\bar{V}_{\mu}\end{array}\right)\left(\begin{array}
{c}\nu_{e_L}\\ \bar\nu_{\mu_R}\end{array}\right).
\ee
Here $V_e$ and $\bar{V}_\mu$ are the matter induced potentials for electron neutrinos 
and muon antineutrinos respectively, and $\mu_{\nu}$ is the neutrino transition 
magnetic moment.

For definiteness we consider $\nu_{e_L}\rightarrow\bar\nu_{{\mu}_R}$ transitions
but all our results apply to $\nu_{e_L}\rightarrow\bar\nu_{{\tau}_R}$ transitions 
as well.
We then proceed with the integration of this system of equations using the solar
density and neutrino fluxes \cite{hom}, thus obtaining the survival and conversion 
probabilities of left handed electron neutrinos. In our numerical calculations we
use $\mu_{\nu}=10^{-11}\mu_{B}$. Since the neutrino magnetic moment enters into the 
evolution equation only in the combination $\mu_{\nu}B$, the results would also
apply to any other value of $\mu_{\nu}$ provided that the solar magnetic field strength 
is rescaled accordingly. For each magnetic field profile the survival probability is 
given in figs. 3  and 4 for the values of $\Delta m^2_{21}$ and $B_0$ corresponding to 
the respective best fit. We note the almost total survival of low energy neutrinos,
the strong suppression of intermediate energy ones and the moderate suppression of the 
most energetic $^8$B ones. The high energy limit is, in the RSFP case, close to
1/2, as confirmed by these figures, a different situation from the small mixing angle 
MSW where this limit is close to unity. This fact leads in general to better fits of 
the rates in the case of the RSFP mechanism than in the MSW one. This is because in the 
case of the MSW mechanism one has to choose very carefully the value of $\Delta
m^2_{21}$ in order to achieve a factor of $\sim$ 1/2 reduction of the high energy
portion of $^8$B neutrinos, whereas this comes out automatically in the case of the 
RSFP mechanism. 

The next step is to insert the survival probability into the expression for the 
event rate in each experiment and take the ratio to the SSM event rate.
The event rate of the gallium experiments in the framework of the RSFP scenario is
calculated as
\be
\bar{R}_{Ga}=\sum_{i}R_{Ga,i}\frac{\bar{R}_{Ga,i}}{R_{Ga,i}}
\ee
where the sum extends over the relevant neutrino fluxes, $(i=pp$, $pep$, $^7$Be,  
$^8$B, $^{13}$N, $^{15}$O). The quantities $\bar{R}_{Ga,i}$ denote the partial
event rate for flux $i$,  as predicted by the RSFP model, where, except for
$i=^7$Be, $pep$ we have
\be
\bar{R}_{Ga,i}=\int_{{E_{i}}_{min}}^{{E_{i}}_{max}}\sigma_{Ga}(E)P(E)
f_{i}(E)dE.
\ee
Here $f_{i}(E)$ represents the $i$-th flux from ref. \cite{hom}, and the integral is
taken from the experimental threshold $E_{min}=E_{th}=0.236$ MeV for the $pp$, $^{15}$O,
$^{13}$N 
and $^8$B cases. For $i=^7$Be, $pep$, the quantities 
$\bar{R}_{Ga,i}$ are just reduced to the survival probability at the corresponding energy.
The quantities $R_{Ga,i}$ are on the other hand the corresponding 
SSM predictions for the partial rates \cite{BP98} obtained from (19) with 
$P=1$ (see also table I).
The RSFP predicted ratio of the rates for the Gallium experiment is finally
\be
r_{Ga}=\frac{\bar{R}_{Ga}}{R_{Ga}}
\ee
where ${R_{Ga}}$ is the SSM total rate for the gallium experiment, 
${R_{Ga}}=\sum_{i}R_{Ga,i}$.

\begin{center} 
\begin{tabular}{cccc} \hline \hline
Source     &    Flux   &   Ga  &  Cl   \\  
           &($10^{10}cm^{-2}s^{-1})$&(SNU)&(SNU)  \\ \hline
pp         & 5.94      & 69.6  & 0.0 \\
pep        & $1.39\times10^{-2}$& 2.8 & 0.2 \\
hep        & $2.1\times10^{-7}$& 0.0 & 0.0 \\
$^7\rm{Be}$& $4.8\times10^{-1}$& 34.4 & 1.15 \\
$^8\rm{B}$ & $5.15\times10^{-4}$& 12.4 & 5.9 \\
$^{13}\rm{N}$& $6.05\times10^{-2}$& 3.7  & 0.1 \\
$^{15}\rm{O}$& $5.32\times10^{-2}$& 6.0  & 0.4 \\
$^{17}\rm{F}$& $6.33\times10^{-4}$& 0.1  & 0.0 \\ \hline
Total      &                    &$129\pm^{8}_{6}$&$7.7\pm^{1.2}_{1.0}$ \\ \hline 
\end{tabular}
\end{center}

{\it{Table I - Fluxes and partial rates ($R_{ji}$ coefficients in the main text) as predicted
by the solar standard model}} \cite{BP98}.

{\vspace{5mm}}

In order to obtain $r_{Cl}$ this procedure is repeated for the chlorine experiment with
the obvious omission of pp neutrinos and the truncation of the other continuous fluxes 
below $E_{th}=0.814$ MeV, the energy threshold in the Cl experiment. The SSM rates
$R_{Cl,i}$ 
are also listed in table I. Regarding SuperKamiokande only one neutrino flux is present,
namely $^8$B 
\footnote{The contribution of a small flux of $hep$ neutrinos is completely 
negligible in the total rates but may be of some importance for the high energy part of 
the recoil electron spectrum in SuperKamiokande. We shall discuss this point in more 
detail in section 4.}, 
and we have for the RSFP predicted ratio
\be
\bar{R}_{SK}=\frac{\int_{E_{min}}^{E_{max}}f_{^8\rm B}(E)\int_{T_{min}}^{T_{max}}
[P(E)\frac{d^2 \sigma_W}{dTdE}+[1-P(E)]\frac{d^2\sigma_{\bar{W}}}{dTdE}]dTdE}
{\int_{E_{min}}^{E_{max}}f_{^8\rm B}(E)\int_{T_{min}}^{T_{max}}\frac{d^2\sigma_{W}}
{dTdE}dTdE}.
\ee 
Here T is the electron recoil energy, $T_{min}=E_{e_{th}}-m_e$ with $E_{e_{th}}=5.5$
MeV \cite{SuperK}
\be
T_{max}=\frac{2E^2}{2E+m_e}
\ee
and the integration limits for the neutrino energy are respectively,
\be
E_{min}=\frac{T_{min}+\sqrt{{T^2}_{min}+2m_{e}T_{min}}}{2}
\ee
and $E_{max}=15$ MeV. The weak interaction differential cross section for 
$\nu_{e} e$ scattering is given by 
\be
\frac{d^2\sigma_W}{dTdE}=\frac{{G_F}^2}{2\pi m_{e}}[(g_V+g_A)^2+
(g_V-g_A)^2\left(1-\frac{T}{E}\right)^2-({g_V}^2-{g_A}^2)\frac{m_{e}T}{E^2}]
\ee
with $g_V=\frac{1}{2}+2sin^2\theta_{W}$, $g_A=\frac{1}{2}$. The cross section for 
$\bar\nu_{\mu} e$ and $\bar\nu_{\tau} e$ scattering is on the other hand
\be
\frac{d^2\sigma_{\bar{W}}}{dTdE}=\frac{{G_F}^2}{2\pi m_{e}}[(g_V-g_A)^2+
(g_V+g_A)^2\left(1-\frac{T}{E}\right)^2-({g_V}^2-{g_A}^2)\frac{m_e T}{E^2}]
\ee
with $g_V=-\frac{1}{2}+2sin^2\theta_{W}$, $g_A=-\frac{1}{2}$.
The contribution of the neutrino electromagnetic cross section to eq. (21) is
negligible, typically 3 orders of magnitude lower than the weak cross section and
will be omitted. We do not include the electron energy resolution function of 
the detector as its effect on the total rate is negligible; we will, however, 
include it in the calculation of the recoil electron spectrum. 
We calculate the ratios $r_{Ga},r_{Cl},\bar{R}_{SK}$ in the parameter ranges
$\Delta m^2_{21}=(4-20)\times 10^{-9}$ eV$^2$ and $B_{0}=(3-30)\times10^4$ G for
all seven magnetic field profiles presented above and identify in each case those 
intervals in which all three ratios simultaneously approach their experimental
values (table II). We then perform for each interval a $\chi^2$ analysis in 
order to evaluate the quality of the fit.

\begin{center}
\begin{tabular}{lcccc} \\ \hline \hline
Experiment &  Data      &   Theory   &   Data/Theory  &  Reference \\ \hline
Homestake  &  $2.56\pm0.16\pm0.15$ & $7.7\pm^{1.2}_{1.0}$ & $0.33\pm0.03$ & \cite{Hom} \\
Gallex     &  $76.4\pm6.3\pm^{4.5}_{4.9}$ & $129\pm ^8_6$ & $0.59\pm0.06$ &
\cite{Gallex} \\
SAGE       &  $69.9\pm^{8.0\pm3.9}_{7.7\pm4.1}$ & $129\pm^8_6$ & $0.54\pm0.06$ & 
\cite{SAGE} \\ SuperKamiokande&$2.44\pm0.05\pm^{0.09}_{0.06}$ & 
$5.15\pm^{1.0}_{0.7}$&$0.474\pm0.020$& \cite{SuperK}\\ \hline
\end{tabular}
\end{center}

{\it{Table II - Data from the four solar neutrino experiments. Units are SNU for 
the first three experiments and in $cm^{-2}s^{-1}$ for SuperKamiokande. For the 
Gallium experiments we have used the combined result $72.3\pm5.6$ {\rm SNU}.}} 


\section{Analysis of the data}

\subsection{Description of $\chi^2$ analysis}

We follow in this subsection the procedure outlined in ref. \cite{FL} adapted to the
RSFP mechanism and the present experimental situation \cite{Hom}-\cite{SuperK}. We 
start with the definition of the $\chi^2$ function
\be
\chi^2=\sum_{j_{1},j_{2}=1}^{3}(\bar{R}_{j_{1}}-{R_{j_{1}}}^{exp})\left[{\sigma^2}
(tot)\right]^{-1}_{j_{1}j_{2}}(\bar{R}_{j_{2}}-{R_{j_{2}}}^{exp}). 
\ee
In this expression the indices $j_{1},j_{2}$ run over the three types of experiments:
Ga, Cl and SuperKamiokande respectively. The quantities $\bar{R}$ denote the theoretical
(RSFP) event rates defined in eqs. (18), (19) for Ga and Cl experiments, and in
(20) for 
the SuperKamiokande one and the quantities $R^{exp}$ denote the
experimental ones given in table II. They are expressed in SNU for
Ga, Cl and in terms of the ratio data/SSM for SuperKamiokande. For the Gallium case we use 
the combined value from SAGE and Gallex, $R^{exp}_{Ga}=72.3\pm5.6$ SNU.
The matrix ${\sigma^2}_{j_{1}j_{2}}(tot)$ is the total error matrix given by
\be
{\sigma^2}_{j_{1}j_{2}}(tot)={\sigma^2}_{j_{1}j_{2}}(exp)+{\sigma^2}_{j_{1}j_{2}}(th).
\ee
The experimental error matrix is given in terms of ${{\sigma}_{j}}^{exp}$, the 
experimental errors shown in table II,
\be
{\sigma^2}_{j_{1}j_{2}}(exp)=\delta_{j_{1}j_{2}}{{\sigma}_{j_{1}}}^{exp}
{{\sigma}_{j_{2}}}^{exp}
\ee
so that
\be
{\sigma^2}_{j_{1}j_{2}}(exp)=\left(\begin{array}{ccc}31.36&0&0\\0&0.0481&0\\
0&0&0.0004\end{array}\right).
\ee
The theoretical error matrix in eq. (27) is the sum of the cross section contribution 
containing the uncertainties in the cross sections with the  
astrophysical contribution involving the uncertainties in the astrophysical parameters 
\be
{\sigma^2}_{j_{1}j_{2}}(th)={\sigma^2}_{j_{1}j_{2}}(cs)+{\sigma^2}_{j_{1}j_{2}}(ap).
\ee
The uncertainties in the detector cross sections, expressed in the form 
$\Delta \ln C_{ij}$,
entering the cross section error matrix, are assumed uncorrelated and were taken from 
ref. \cite {BP92}. They give rise to the following diagonal matrix
\be
{\sigma}^2_{j_{1}j_{2}}(cs)=\delta_{j_{1}j_{2}}\sum_{i=1}^6 \bar{R^2}_{ij_{1}}
(\Delta \ln C_{ij_{1}})^2.
\ee
where the index $i$ runs over all relevant fluxes, $i=pp$, $pep$, $^7$Be, $^8$B,
$^{13}$N, $^{15}$O. Model dependence appears in this matrix through the 
$\bar{R}_{ij}'s$.

The astrophysical error matrix is finally
\be
{\sigma^2}_{j_{1}j_{2}}(ap)=\sum_{i_{1},i_{2}=1}^{6}\bar{R}_{j_{1}i_{1}}\bar{R}_{i_{2}j_{2}}
\sum_{k=1}^{9}
\alpha_{i_{1}k}\alpha_{i_{2}k}(\Delta \ln X_{k})^2
\ee
In this expression index k extends over the nine astrophysical parameters
$k=S_{11},S_{33},S_{34},\\S_{1,14},S_{17},Lum,Z/X,Age,Opac$, the ${\alpha}$ matrix denotes 
the logarithmic 
derivatives of the six fluxes considered with respect to these parameters, 
${\alpha}_{ik}=
{\partial}\ln f_{i}/{\partial}\ln X_{k}$ 
and the rest of the notation is self clear. The $(6\times9)$ matrix
\be
\alpha_{ik}\, \Delta\ln X_k 
\ee
which appears in equation (32) was calculated using the data from \cite{hom}. 
We now have all the necessary elements to
calculate $\chi^2$ for each magnetic field profile.

\subsection{Fits of the rates and their quality}

In this subsection we will investigate the local minima of $\chi^2$ for all studied solar field
profiles, identifying in each profile the lowest minimum in terms of $B_0$ (peak field value)
and $\Delta m^2_{21}$. As discussed in the introduction we analyze the intervals
$B_{0}=(3-30)\times10^4$ G and $\Delta m^2_{21}=(4-20)\times10^{-9}$ eV$^2$, the
first interval dictated 
by the requirement of significant neutrino conversion (lower bound) and solar physics arguments
\cite{Parker} (upper bound) and the second interval dictated by the resonance locations of the 
several neutrino fluxes. In fitting the rates we have three types of experiments (Ga, Cl, 
SuperKamiokande)
and two parameters: $\Delta m^2_{21}$, $\mu_{\nu}B_0$ for a given magnetic field profile. Hence the number
of degrees of freedom is one. We  
search for values of $\chi^2$ smaller than or of order unity and discuss the profiles in the same
order as in section 2. Starting therefore with profile 1 (equilateral triangle), 
eqs. (1)-(3), we observe two local minima (fig. 5), one at $B_0\simeq1.68\times10^5$ G 
and the other at $B_0\simeq2.71\times10^5$ G, both with $\Delta m^2_{21}\simeq 8
\times10^{-9}$eV$^2$. The first 
is an absolute minimum with ${\chi^2}_{min}=0.085$ and for the second, 
${\chi^2}_{min}=0.137$. While such values of $\chi^2$ (especially the first one) are 
remarkably low, the 
stability of both fits against small parameter variations is rather poor: a change of 
3 kG in the value of $B_0$ implies a change in $\chi^2$ by a factor of order 30.    

The second profile is the triangle field, eqs (1)-(3), in which we identified three 
local minima (fig. 6):
the best fit (${\chi^2}_{min}=0.10$) corresponds to $B_0\simeq1.23\times10^5$ G,
$\Delta m^2_{21}
\simeq1.20\times10^{-8}$ eV$^2$, the second best (${\chi^2}_{min}=0.284)$ to
$B_0\simeq1.98
\times10^5$ G, $\Delta m^2_{21}\simeq1.20\times10^{-8}$ eV$^2$ and the third
(${\chi^2}_{min}=0.553$)
to $B_0\simeq2.85\times10^{5}$ G, $\Delta m^2_{21}\simeq1.25\times10^{-8}$ eV$^2$.
These minima are
displayed in fig. 6 and it is seen that their stability is better than the previous
cases: a change of 3 kG implies a change in $\chi^2$ by a factor of 3-4.   

Next we examine profile 3, eqs. (4)-(6): there are two local minima of $\chi^2$ 
both corresponding  
to $\Delta m^2_{21}\simeq1.2\times10^{-8}$ eV$^2$ (fig. 7), the first one at
$B_0=9.54\times10^4$ G with an 
impressively low $\chi^2$ $(\chi^2_{min}=0.036)$ and the second one at 
$B_0\simeq1.72\times10^5$ G
(${\chi^2}_{min}=0.11$). We note the extreme instability of the first, where the same change of
3 kG in $B_0$ leads to a change in $\chi^2$ by a factor of 50, while for the second
the corresponding factor is of order 30.

Profile 4, eqs. (7), (8), shows two local minima of 
$\chi^2$ (fig. 8): ${\chi^2}_{min}=0.68$ for $\Delta
m^2_{21}\simeq1.2\times10^{-8}$ eV$^2$,
$B_0=9.0\times10^4$ G and ${\chi^2}_{min}=0.59$ for $\Delta m^2_{21}\simeq 1.6
\times10^{-8}$ eV$^2$,
$B_0=1.7\times10^5$ G. These fits are poorer than the previous ones and in addition
they are also 
quite unstable: a factor of at least 10 increase in $\chi^2$ for a 3 kG change in
the field at the peak. 
Of all cases examined, such a profile appears to be the least favoured by the data. If 
in this case we choose a power greater than 2 in eq. (8), the results become even
worse as the SuperKamiokande and Chlorine rates become too close to each other.

The essential difference between profile 5, eqs. (9), (10), and the previous one
is the shape of the decrease along the convective zone. This is enough to ensure a 
totally different quality of the fit. In fact we observe only one very stable
minimum in the whole $B_0$ range with an extremely low ${\chi^2}$ (fig. 9):
${\chi^2}_{min}=0.054$ for $\Delta m^2_{21}\simeq
2.1\times10^{-8}$ eV$^2$, $B_0\simeq1.45\times10^5$ G. The stability of this
minimum is in sharp contrast with previous cases, as a change of 15 kG in the peak
field implies a change in ${\chi^2}$ by a factor of 2 at the most. An obvious conclusion 
from the comparison with the previous case is that the data prefers a profile with an
upward facing concavity relative to a downward one along the convective zone. This is to 
be expected, since the low energy
sector of the $^8$B neutrinos necessarily undergoes a strong suppression as the
corresponding energy
is in the intermediate range, where suppression is maximal. Hence in order to ensure a 
moderate suppression for the whole $^8$B flux, its high energy part must be
suppressed to a much lesser extent implying
a strong decrease of the field as from the start of the convective zone. It should be 
noted that such profiles are also expected to provide better fits of the high energy
part of the recoil electron spectrum in SuperKamiokande, as they lead to a weaker
suppression of the high energy part of the solar neutrino spectrum (see next
subsection).  

Profile 6, eqs. (11)-(13), also shows a remarkably good and stable fit. Its general
shape is roughly the same as the previous one, the only difference being a finite upward slope at the 
bottom of the convective zone. The best fit (${\chi^2}_{min}=0.047$) is for $\Delta m^2_{21}\simeq
1.60\times10^{-8}$ eV$^2$, $B_0\simeq9.6\times10^4$ G (see fig. 10). It is also
seen that a change of
11 kG implies a change by a factor of 5 in $\chi^2$. A second, poorer fit, also
shows up in this
case: ${\chi^2}_{min}=0.65$ for $\Delta m^2_{21}\simeq1.3\times10^{-8}$ eV$^2$,
$B_0\simeq 2.63\times10^5$ G.

Finally we examined profile 7. There are four local minima of $\chi^2$ in this case, all
providing fits that are poor and relatively unstable in comparison with the previous ones. We have
(see fig. 11) : ${\chi^2}_{min}=0.84$ for $\Delta
m^2_{21}\simeq1.1\times10^{-8}$ eV$^2$,
$B_0\simeq1.26\times10^5$ G, ${\chi^2}_{min}=0.53$ for $\Delta
m^2_{21}\simeq1.2\times10^{-8}$ eV$^2$,
$B_0\simeq1.66\times10^5$ G, ${\chi^2}_{min}=0.71$ for $\Delta
m^2_{21}\simeq1.4\times10^{-8}$ eV$^2$,
$B_0\simeq2.3\times10^5$ G, ${\chi^2}_{min}=0.875$ for $\Delta
m^2_{21}\simeq1.2\times10^{-8}$ eV$^2$,
$B_0\simeq2.7\times10^5$ G. A change of 3 kG in the central value of $B_0$ implies
an increase of
$\chi^2$ by a factor of at least 6. This profile is therefore manifestly disfavoured 
relative to profile 6. It is worth noting that the difference between the two lies
only in the shape of the downward slope along the convective zone.      


\subsection{Fits of the recoil spectra and their quality}

In this subsection we calculate the spectrum of recoil electrons in SuperKamiokande for each of 
the magnetic field profiles 1-7.  
We take in each case the values of $(\Delta m^2_{21}, 
B_0)$ that correspond to the best fit of the rates as calculated in the previous subsection
and evaluate the corresponding $\chi^2$ of the spectrum. We also determine
for the cases whose fits and stability are the best (profiles 5 and 6) the minimum $\chi^2$ and 
corresponding $(\Delta m^2_{21}, B_0)$. The starting point is the ratio of the SuperKamiokande 
event rate to the one predicted by the SSM evaluated for the 18 energy bins
($j=1,...,18$)
\be
R_j=\frac
{\int_{E_j}^{E_{j+1}} dE_e \int_0^\infty dE_e'\,f(E_e,E_e')\int_{E_{min}(E'_e)}^{\infty}
dE\,f_{^8\rm B}(E)\left(P(E)\frac{d^2\sigma_{W}}{dTdE}+
[1-P(E)]\frac{d^2\sigma_{\bar{W}}}{dTdE}\right)}
{\int_{E_j}^{E_{j+1}} dE_e \int_0^\infty dE_e'\,f(E_e,E_e')\int_{E_{min}(E'_e)}^{\infty}
dE\,f_{^8\rm B}(E)\,\frac{d^2\sigma_{W}}{dTdE}}
\ee
where, $f(E_e,E_e')$ is the energy resolution function which can be found in \cite{SK},  
$E_e=T+m_e$, and the rest of the notation was defined in section 2. In fig. 12
we plot $R_j$ corresponding to profiles 5 and 6 superimposed on data set II.

As far as the $\chi^2$ analysis is concerned there is only one flux now and the values to be 
fitted are the 18 data points given by the SuperKamiokande collaboration (set I 
\cite{CapeTown} and set II \cite{Ringberg} with 708 days and 825 days respectively of 
data taking). Hence the definition of $\chi^2$ is now
\be
\chi^2=\sum_{j_{1},j_{2}=1}^{18}(\bar{R}_{j_1}-R^{exp}_{j_2})[\sigma^2(tot)]^{-1}_{j_{1}j_{2}}
(\bar{R}_{j_2}-R^{exp}_{j_2})\,.
\ee
Here the quantities $\bar{R}_{j_{1}},\bar{R}_{j_{2}}$ denote the vector elements given by (34). 
Since only one neutrino flux is considered here, namely $^8$B, 
in calculating $\sigma^2(ap)$ the matrix within the second summation in equation (32) has 
now been reduced to its column vector corresponding to $i_1=i_2=4=^8$B. We thus
have 
\be
\sigma^2_{{j_1}{j_2}}(ap)=\bar{R}_{j_{1}}\bar{R}_{j_{2}}\sum_{k=1}^9\alpha^2_{4k}
(\Delta \ln X_k)^2
\ee
where $j_{1},j_{2}=1,...,18$.
The matrix elements of $\sigma(exp)$ (28) are directly read 
from \cite{CapeTown} and \cite{Ringberg}, the matrix $\sigma(cs)$ can be taken to be 
zero since the uncertainties in the cross sections of $\nu e$ scattering are negligible. 

Evaluating $\chi^2$ for the spectrum using the values of $(\Delta m^2_{21},B_0)$ that correspond to 
the best fit of the rates, we find the results shown in table III for each field profile and data
set. It is worth noting 
that the profiles that generate the best and most stable fits (5 and 6) for the
rates are also those that lead to the lowest $\chi^2$ for the spectrum. We have calculated for these
two the minimum $\chi^2$ (for data set II) and found for case 5 a minimum at
$(3.2\times10^{-8}$ eV$^2, 
1.14\times10^5G)$ $({\chi^2_{sp}}_{min}=23.235)$ to be compared
with (see section 3.2) $(2.1\times10^{-8}$ eV$^2, 1.45\times10^5$ G) for the rates.
For case 6 (data set 
II) we found a minimum at $(2.4\times10^{-8}$ eV$^2, 1.12\times10^5G)$
$({\chi^2_{sp}}_{min}=23.236)$ 
to be compared with (see section 3.2) 
$(1.6\times10^{-8}$ eV$^2, 
9.6\times10^4$ G) for the rates. The stability of these fits should be stressed, as
in fact
the change in the value of $\chi^2$ between the two minima (for the rates and spectrum) in each of 
the two profiles only affects the third digit of $\chi^2$. For this reason we do not include the 
$\chi_{sp}^2$ plots in the paper. 

We note finally that although the fits for data set II are slightly worse than those for data 
set I, the differences in $\chi^2$ are only marginal. Furthermore, one must keep in mind that the 
large uncertainties still present in the high energy $(E_{e}>13$ MeV) part of the
spectrum and the 
differences in the central data points for sets I and II make it difficult to draw any definitive 
conclusions about the particle-physics solution of the solar neutrino problem favoured by 
the recoil electron spectrum. 

\begin{center}
\begin{tabular}{ccc} \hline \hline
Profile&  $\chi^2_{sp}(I)$ & $\chi^2_{sp}(II)$ \\ \hline
1      &  22.5 & 24.8 \\ 
2      &  21.9 & 23.9 \\ 
3      &  22.5 & 24.8 \\
4      &  25.9 & 29.5 \\ 
5      &  21.6 & 23.5 \\ 
6      &  21.7 & 23.6 \\  
7      &  22.3 & 24.5 \\ \hline
\end{tabular}
\end{center}
{\it{Table III - The values of $\chi^2$ (16 d.o.f.) for the recoil electron spectrum in 
SuperKamiokande evaluated
for $\Delta m^2_{21}, B_0$ which correspond to the best fits of the rates. Data sets 
(I) and (II) are given in refs. \cite{CapeTown}, \cite{Ringberg} and profiles 1-7 are 
shown in figs. 1 and 2. Cases 5 and 6, besides providing the lowest $\chi^2_{sp}$'s,
also give the lowest $\chi^2$'s for the rates and, interestingly enough, the most stable 
fits of all.}}      


\section{Summary and Discussion}

We have examined the status of the RSFP solution to the solar neutrino
problem in the light of the most recent experimental data on total rates and energy spectrum and the
standard solar model of BP98 \cite{BP98}. There
is independent, although scarce, input information from solar physics on the solar magnetic field
\cite{Parker} suggesting a field that is largest around the bottom of the convective zone with a 
peak up to $3\times10^5$ G in that region. So at the start of any RSFP analysis one
cannot consider
the field profile as entirely free. On the other hand previous studies indicate
that the common suppression of intermediate energy neutrinos and the almost total survival of low
energy ones must lead, if neutrinos have a magnetic moment, to a profile rising sharply over a 
relatively short distance along the solar radius, decreasing then smoothly towards the surface.
Using this input information we investigated a general class of profiles with the above
characteristics and calculated for each of them the values of $\Delta m^2_{21}, B_0$ 
which provide the best fits for the total rates.  

The shape of the survival probability as a function of neutrino energy is obviously 
determinant for the quality of these fits. One important general feature is that the
limit of this probability as the energy increases is close to 1/2 in the case of
RSFP, whereas in the case of the small mixing angle MSW solution, which gives the best 
fit for the rates among the neutrino oscillations solutions, it is close to unity.
Hence the important contribution of the $^8$B 
neutrinos, which survive up to a factor of almost 1/2 and whose energy spread is large, 
is naturally much better fitted in RSFP than in MSW. Thus the fact that the values of 
$\chi^2_{min}$ for the rates best fits range from 0.036 to 0.59 for 1 d.o.f.
(profiles 3 and 4 respectively), to be compared with 1.7 (SMA solution) and 4.3 
(LMA and vacuum oscillation solutions) \cite{BKS}, is not surprising 
\footnote{It should be noted that smaller values of $\chi^2_{min}$ for the  
MSW solutions were obtained in \cite{GGHPV}: $\chi^2_{min}=0.44$ for the SMA and 
$\chi^2_{min}=2.7$ for the LMA solution.}. 

The investigation performed for the rates fits has taught us that, besides a fast rise 
across the bottom of the convective zone, the magnitude of the field should decrease
faster past this bottom 
and then slower on approaching the solar surface. In other words, the profile must show 
an upward facing concavity. To this end it suffices to compare the fits 
of profiles 4 with 5 on one hand and the fits of profiles 6 with 7 on the other. The 
very low $\chi^2$ obtained for profile 3 (0.036), with the reverse concavity, is
extremely unstable as compared to the ``best'' cases 5 and 6. This fact may be easily 
understood if one notes that the low energy sector of the $^8$B neutrinos has to be 
highly depleted because their energies are close to those of $^7$Be, CNO and $pep$ 
neutrinos which must be strongly suppressed by RSFP in order to fit the data. On the 
other hand, the overall suppression of $^8$B neutrinos is just moderate. So the solar 
field must ensure a small suppression of their high energy sector. This is achieved if 
the field becomes much weaker soon after the intermediate energy resonance densities. 
Hence an upward facing concavity of the field profile along the convective zone appears 
to be favoured. 

We have also calculated the recoil electron energy spectra in SuperKamiokande  
and performed the corresponding fits for the same class of field profiles. We found for 
all of them except one (profile 4) the correct sign slope against the recoil
electron energy. Regarding the high energy bins, the central data points show the
excess of the observed number of events compared with the theoretical predictions, but 
it should be stressed that the uncertainties are much larger in this sector. Hence the 
general quality of the fits for the spectra is also encouraging. We found, with 
16 d. o. f., $(\chi^2_{sp})_{min}$ ranging from 21.6 to 25.9 (data set I \cite 
{CapeTown}, profiles 5 and 4 respectively) and 23.4 to 29.5  
(data set II \cite{Ringberg}, also profiles 5 and 4 respectively). 
All minima are quite shallow. These values are to be compared with 25.0
\cite{Ringberg} (24.1 \cite{GGHPV}) for the SMA solution, 23.5 \cite{GGHPV} for the 
LMA solution and 17.4 \cite{Ringberg} for the vacuum oscillation solution 
\footnote{The fact that the recoil electron spectrum (including the observed excess
of the high energy events) can be described reasonably well by vacuum oscillations
was first noted in \cite{SuperK} and the corresponding solutions were found
in \cite{BFL}.}. 
Future data on the SuperKamiokande energy spectrum are thus eagerly awaited, 
especially regarding the improvement in accuracy of its upper and lower ranges. In 
fact, determining whether the spectrum really rises with energy for $E_{e}>13$ MeV 
or else the apparent rise is just a statistical fluctuation or an instrumental effect 
will be essential to discriminate between various solutions of the solar neutrino 
problem. As it stands, no theoretical model (oscillations or RSFP) is able to
provide a very good fit. 

It should be noted that reasonably good fits of the spectrum can be achieved in all 
scenarios if one treats the poorly known flux of $hep$ neutrinos, which constitutes the 
highest energy component of 
the solar neutrino spectrum, as a free parameter \cite{free,BKS2,GGHPV,Ringberg}. We
did not perform such an analysis because the latest SuperKamiokande data shows the 
decreased excess of high energy events \cite{Ringberg} and it may therefore be 
advisable to wait until more accurate data has been obtained. 

We have used throughout a value of $10^{-11}\mu_B$ for the neutrino magnetic moment. 
Since the order parameter is the product $\mu_{\nu}B$ (eq. (17)) and the peak field may 
be as large as $3\times10^5$ G, a
factor of 2-3 larger than its best fits, the appropriate order of magnitude for 
$\mu_{\nu}$ is in fact $(3-5)\times10^{-12}\mu_B$, a value consistent with most
astrophysical bounds \cite{magmo}.

To conclude, the prospects of RSFP as a possible solution to the solar neutrino problem 
are at present quite encouraging: in fact the fits to the total rates are excellent,
far better than in any other scenario \cite{BKS,BKS2,GGHPV}. As far as the fits for the 
recoil electron energy spectrum are concerned, they are in general slightly better than 
the small mixing angle MSW ones, similar to the large mixing angle MSW ones and worse 
than vacuum oscillation ones, depending of course on the appropriate choice of the field 
profile. The improvement in the data on the spectrum, will be essential in helping to 
ascertain the most realistic theoretical model.

\section*{Acknowledgements}

The work of one of us (E. A.) work was supported by Funda\c{c}\~ao para a Ci\^encia e 
a Tecnologia through the grant PRAXIS XXI/BCC/16414/98 and also in part by the 
TMR network grant ERBFMRX-CT960090 of the European Union. We are grateful to 
M. Guzzo and H. Nunokawa for useful correspondence.

\newpage

\centerline{\large Figure captions}

\vglue 0.4cm
\noindent
Fig. 1.
Solar magnetic fields as functions of the solar coordinate 
$x=r/R_{S}$. Field strength is in Gauss. Solid line is profile 1 [eqs. (1) - (3)
with  $x_R=0.7, x_C=0.85$], dotted line is profile 2 [eqs. (1) - (3) with $x_R=0.65,
x_C=0.8$], dashed line is profile 3 [eqs. (4) - (6)], long-dashed line is profile 
4 [eqs. (7), (8)]. For $x>0.75$ profiles 3 and 4 coincide. 

\noindent
Fig. 2.
Profile 5 [solid line, eqs. (9), (10)] profile 6 [dotted line, eqs. (11) -(13)],
profile 7 [dashed line, eqs.(14) - (16)] . For $0.71\leq x\leq 0.75$ profiles 6 and
7 coincide. 

\noindent
Fig. 3.
Neutrino survival probabilities for profiles 1 - 4 as functions of
neutrino energy for $\Delta m^2_{21}$ and $B_0$ given by their values at the best
fit of the rates for each profile. Solid line corresponds to profile 1, dotted line,  
to profile 2, dashed line, to profile 3, long dashed line, to profile 4. 

\noindent
Fig. 4.
Same as fig. 3 for the neutrino survival probabilities for profiles 5 - 7.
Solid line corresponds to profile 5, dotted line to profile 6 and dashed line to
profile 7. 

\noindent
Fig. 5.
Rate fits: values of $\chi^2/d.o.f.$ as functions of the peak value of the 
field for profile 1 for several values of $\Delta m^2_{21}$. Left: dashed line
corresponds to $7\times 10^{-9}$ eV$^2$, full to $8\times10^{-9}$ eV$^2$, dotted to
$9\times10^{-9}$ eV$^2$.Right: dotted line corresponds to $7\times10^{-9}$ eV$^2$, 
full to $8\times10^{-9}$ eV$^2$, dashed to $9\times10^{-9}$ eV$^2$. Best fit:
$\Delta m^2_{21}=8\times10^{-9}$ eV$^2$, $B_0=1.68\times 10^{5}$ G, 
$\chi^2_{min}/d.o.f.=0.085$.

\noindent
Fig. 6.
Rate fits: $\chi^2/d.o.f.$ as a function of the peak field value for 
several values of $\Delta m^2_{21}$ for profile 2. Top left: dashed $1.15 
\times10^{-8}$ eV$^2$, full $1.2\times10^{-8}$ eV$^2$, dotted $1.25\times10^{-8}$
eV$^2$. Top right: full $1.3\times10^{-8}$ eV$^2$, dotted $1.35\times10^{-8}$ eV$^2$, 
dashed $1.40\times10^{-8}$ eV$^2$. Bottom left: dotted $1.15\times10^{-8}$ eV$^2$,
full $1.20\times10^{-8}$ eV$^2$, dashed $1.25 \times10^{-8}$ eV$^2$. Bottom right:
dotted $1.15\times10^{-8}$ eV$^2$, dashed $1.20\times 10^{-8}$ eV$^2$, full
$1.25\times10^{-8}$ eV$^2$. Best fit: $\Delta m^2_{21}=1.20\times10^{-8}$ eV$^2$, 
$B_0=1.23\times10^{5}$, $\chi^2_{min}/d.o.f.=0.100$.

\noindent
Fig. 7.
Rate fits: $\chi^2/d.o.f.$ as a function of the peak field (profile 3) for 
several values of $\Delta m^2_{21}$. Left: dotted $1.1\times10^{-8}$ eV$^2$, full
$1.2\times10^{-8}$ eV$^2$, dashed $1.3\times10^{-8}$ eV$^2$. Right: dot-dashed
$1.1\times10^{-8}$ eV$^2$, full $1.2\times10^{-8}$ eV$^2$, dotted $1.3\times10^{-8}$ 
eV$^2$, dashed $1.4\times10^{-8}$ eV$^2$. Best fit: $\Delta m^2_{21}=1.2\times10^{-8}$ 
eV$^2$, $B_{0}=9.54\times10^{4}$ G, $\chi^2_{min}/d.o.f.=0.036$.

\noindent
Fig. 8.
Rate fits: $\chi^2/d.o.f.$ as a function of the peak field (profile 4) for 
several values of $\Delta m^2_{21}$. Left: dashed $1.1\times10^{-8}$ eV$^2$, full
$1.2\times10^{-8}$ eV$^2$, dotted $1.3\times10^{-8}$ eV$^2$. Right: dashed
$1.5\times10^{-8}$ eV$^2$, full $1.6 \times10^{-8}$ eV$^2$, dotted
$1.7\times10^{-8}$ eV$^2$. Best fit: $\Delta m^2_{21}=1.6\times10^{-8}$ eV$^2$, 
$B_{0}=1.7\times10^{5}$G, $\chi^2_{min}/d.o.f.=0.59$.

\noindent
Fig. 9.
Rate fits: $\chi^2/d.o.f.$ as a function of the peak field for profile 5 for
several values of $\Delta m^2_{21}$. Dotted line $1.9\times10^{-8}$ eV$^2$, dashed
$2.0\times10^{-8}$ eV$^2$, full $2.1\times10^{-8}$ eV$^2$, dot-dashed 
$2.2\times10^{-8}$ eV$^2$. Best fit: $\Delta m^2_{21}=2.1\times10^{-8}$ eV$^2$, 
$B_{0}=1.45\times10^{5}$ G, $\chi^2_{min}/d.o.f.=0.0547$.

\noindent
Fig. 10.
Rate fits: $\chi^2/d.o.f.$ as a function of the peak field for several 
values of $\Delta m^2_{21}$ (profile 6). Left: dotted $1.55\times10^{-8}$ eV$^2$, 
full $1.6\times 10^{-8}$ eV$^2$, short dashed $1.65\times10^{-8}$ eV$^2$, long
dashed $1.70\times10^{-8}$ eV$^2$, dot-dashed $1.75\times10^{-8}$ eV$^2$. Right:
dotted $1.2\times10^{-8}$ eV$^2$, full $1.3\times10^{-8}$ eV$^2$, dashed
$1.4\times10^{-8}$ eV$^2$. Best fit: $\Delta m^2_{21}=1.6\times10^{-8}$ eV$^2$, 
$B_0=9.6\times10^{4}$ G, $\chi^2_{min}/d.o.f.=0.0477$.

\noindent
Fig. 11.
Rate fits: $\chi^2/d.o.f.$ as a function of the peak field (profile 7) for 
several values of $\Delta m^2_{21}$. Top left: dotted $1.0\times10^{-8}$ eV$^2$, full
$1.1\times10^{-8}$ eV$^2$, dashed $1.2\times10^{-8}$ eV$^2$, dot-dashed
$1.3\times10^{-8}$ eV$^2$. Top right: dotted $1.1\times10^{-8}$ eV$^2$, full
$1.2\times10^{-8}eV^2$, dashed $1.3\times10^{-8}$ eV$^2$.
Bottom left: dotted $1.0\times10^{-8}$ eV$^2$, short dashed $1.1\times10^{-8}$ 
eV$^2$, long dashed $1.2\times10^{-8}$ eV$^2$, dot-dashed $1.3\times10^{-8}$
eV$^2$, full $1.4\times10^{-8}$ eV$^2$. Bottom right: dotted
$1.0\times10^{-8}$ eV$^2$, dot-dashed $1.1\times10^{-8}$ eV$^2$,
full $1.2\times10^{-8}$ eV$^2$, short dashed $1.3\times10^{-8}$ eV$^2$. Best fit:
$\Delta m^2_{21}=1.2\times 10^{-8}$ eV$^2$, $B_{0}=1.66\times10^{5}$ G,
$\chi^2_{min}/d.o.f.=0.53$. 

\noindent
Fig. 12.
Spectrum fits: for profiles 5 (top) and 6 (bottom) we show the predicted 
electron recoil spectrum as a function of total electron energy superimposed on data
set (II) (825 days), ref. \cite{Ringberg}, with $\Delta m^2_{21}$ and $B_0$ given by
their values corresponding to the rates best fits (see figs. 9 and 10). 
For profile 5 $\chi^2_{sp}/d.o.f.=23.5/16$, and for profile 6
$\chi^2_{sp}/d.o.f.=23.6/16$. For both $\chi^2_{{sp}_{min}}/d.o.f.=23.2/16$ with
values of $\Delta m^2_{21}, B_{0}$ that are slightly different from the best 
fit ones for the rates (see the main text).

\newpage

\begin{figure}
\mbox{\psfig{figure=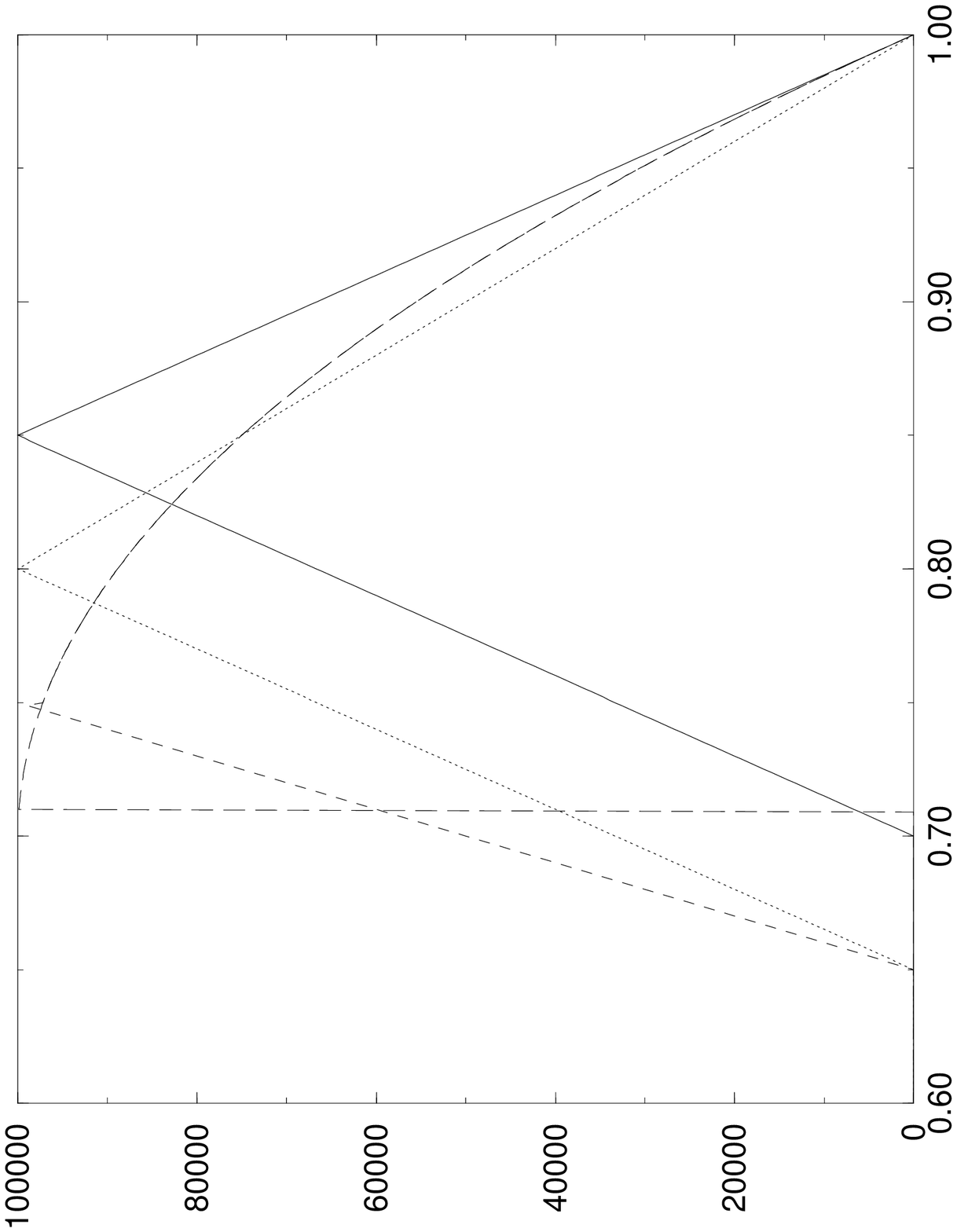,width=17cm}}
\centerline{\mbox{Fig. 1.}}
\end{figure}
\newpage

\begin{figure}
\mbox{\psfig{figure=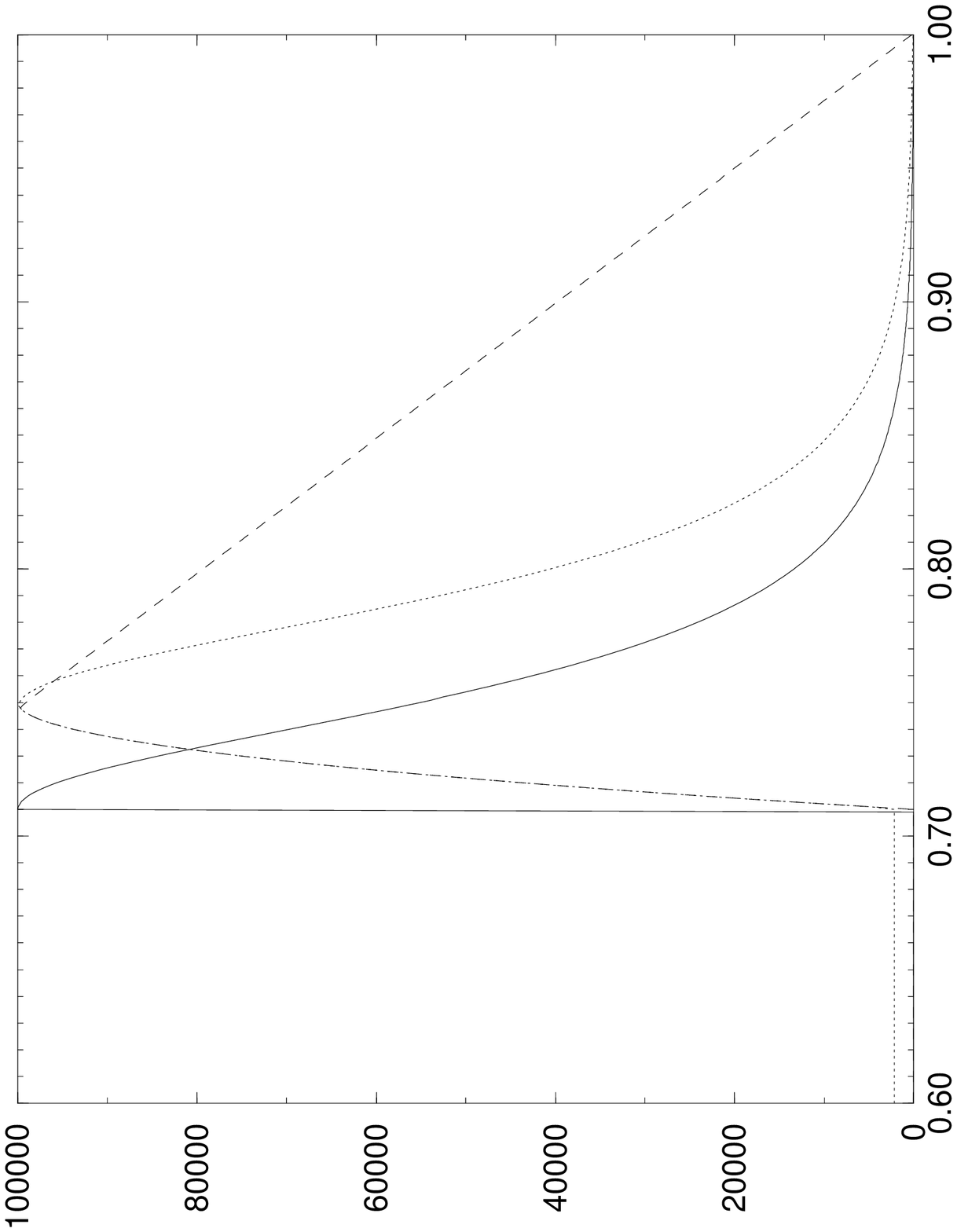,width=17cm}}
\centerline{\mbox{Fig. 2.}}
\end{figure}
\newpage

\begin{figure}
\mbox{\psfig{figure=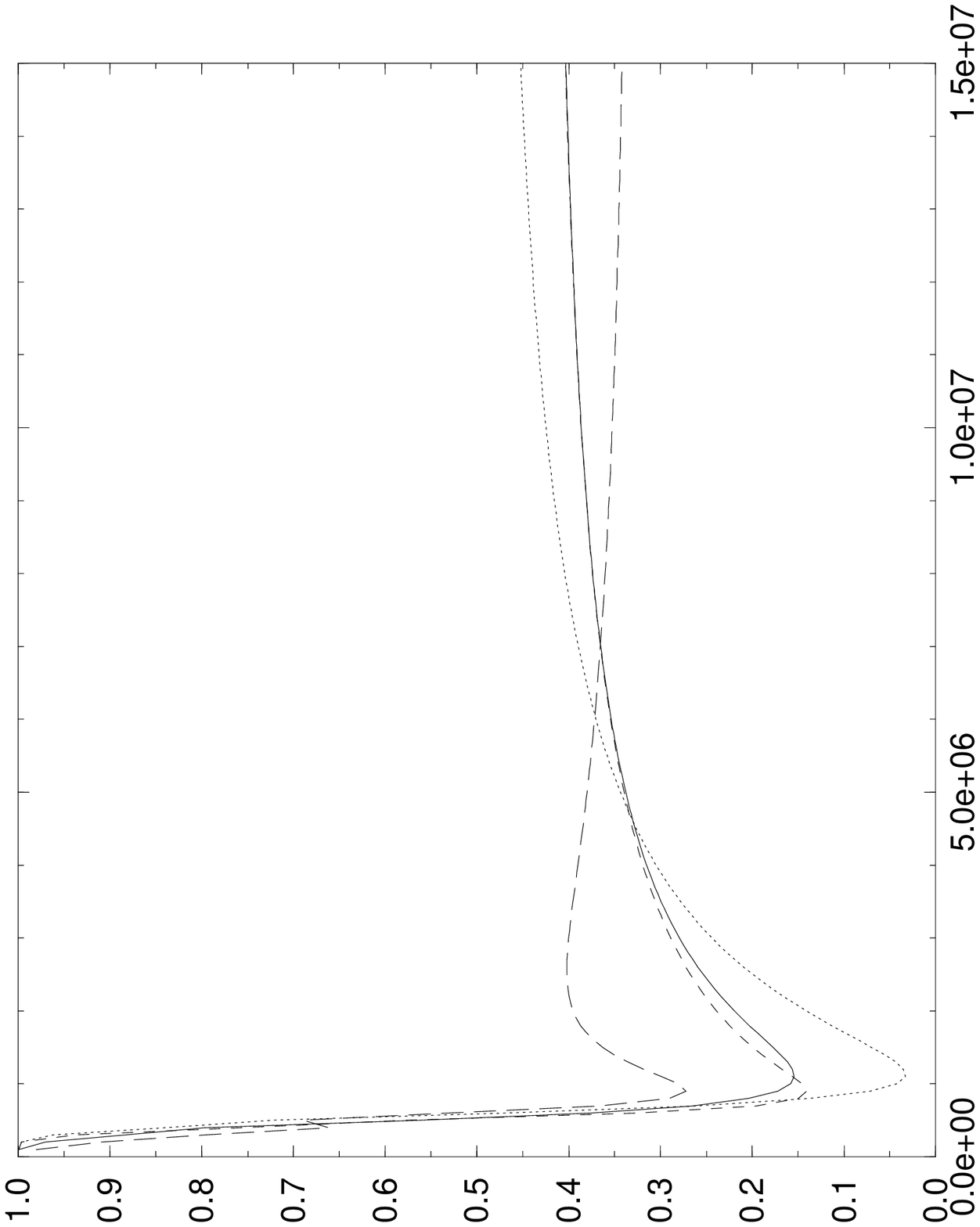,width=15cm}}
\centerline{\mbox{Fig. 3.}}
\end{figure}

\begin{figure}
\mbox{\psfig{figure=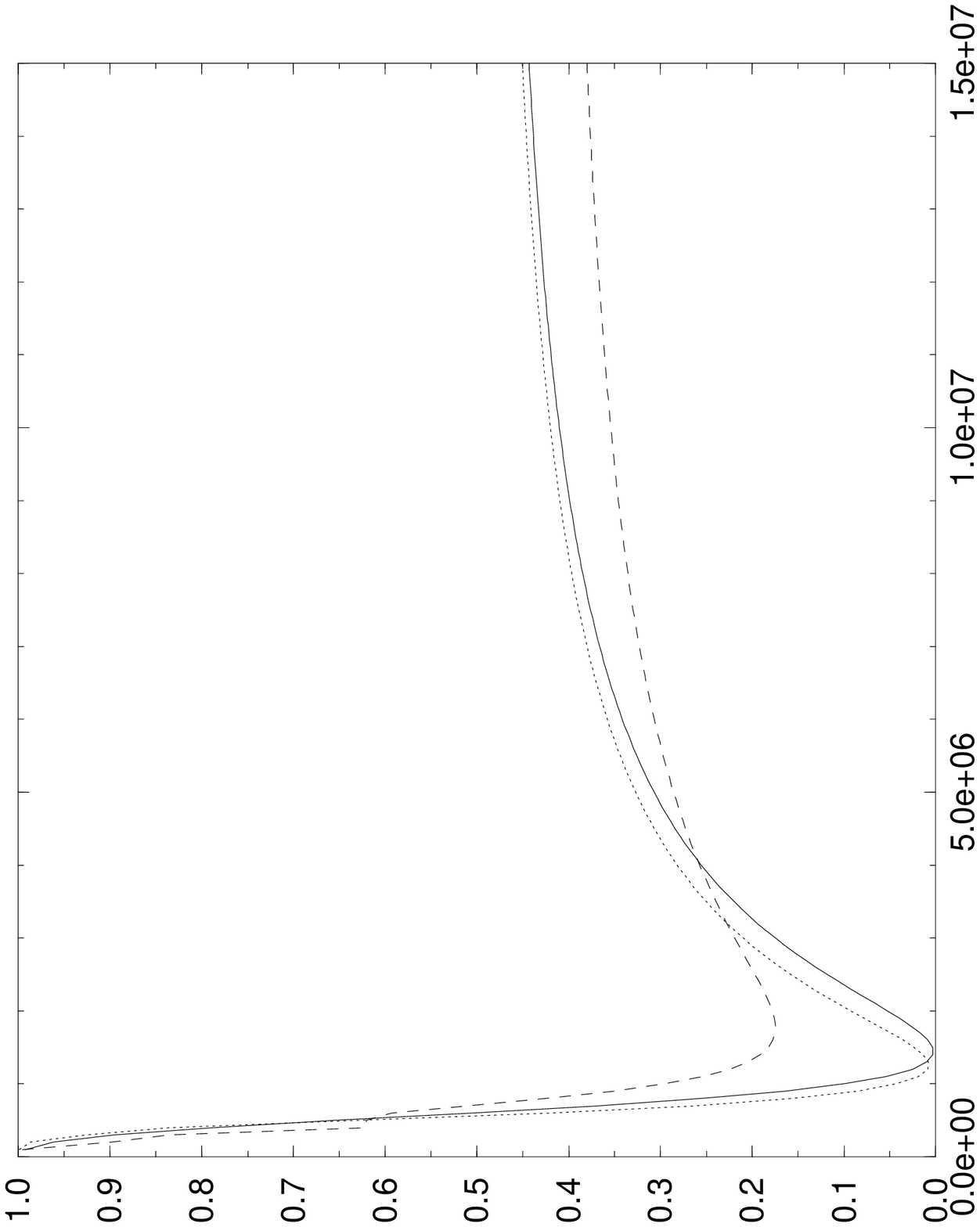,width=15cm}}
\centerline{\mbox{Fig. 4.}}
\end{figure}

\begin{figure}
\mbox{\psfig{figure=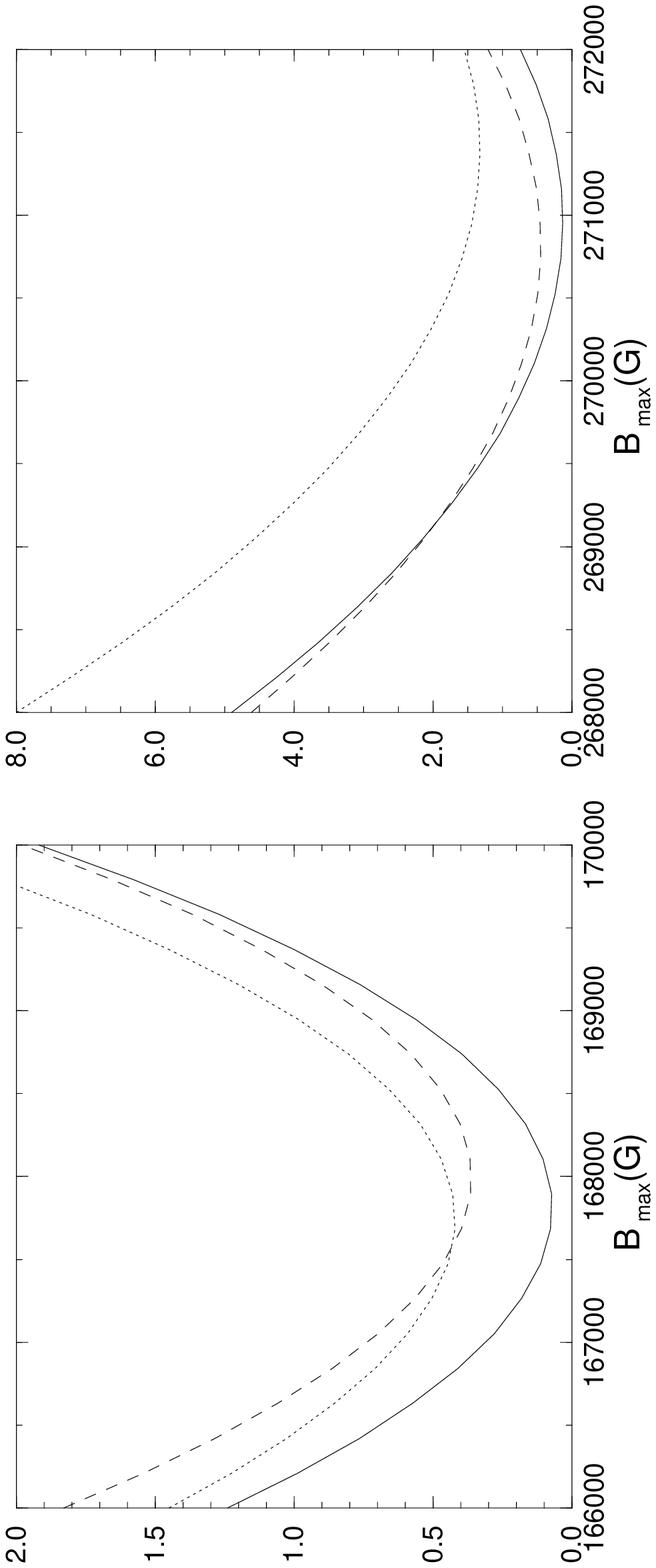,width=15cm}}
\centerline{\mbox{Fig. 5.}}
\end{figure}

\begin{figure}
\mbox{\psfig{figure=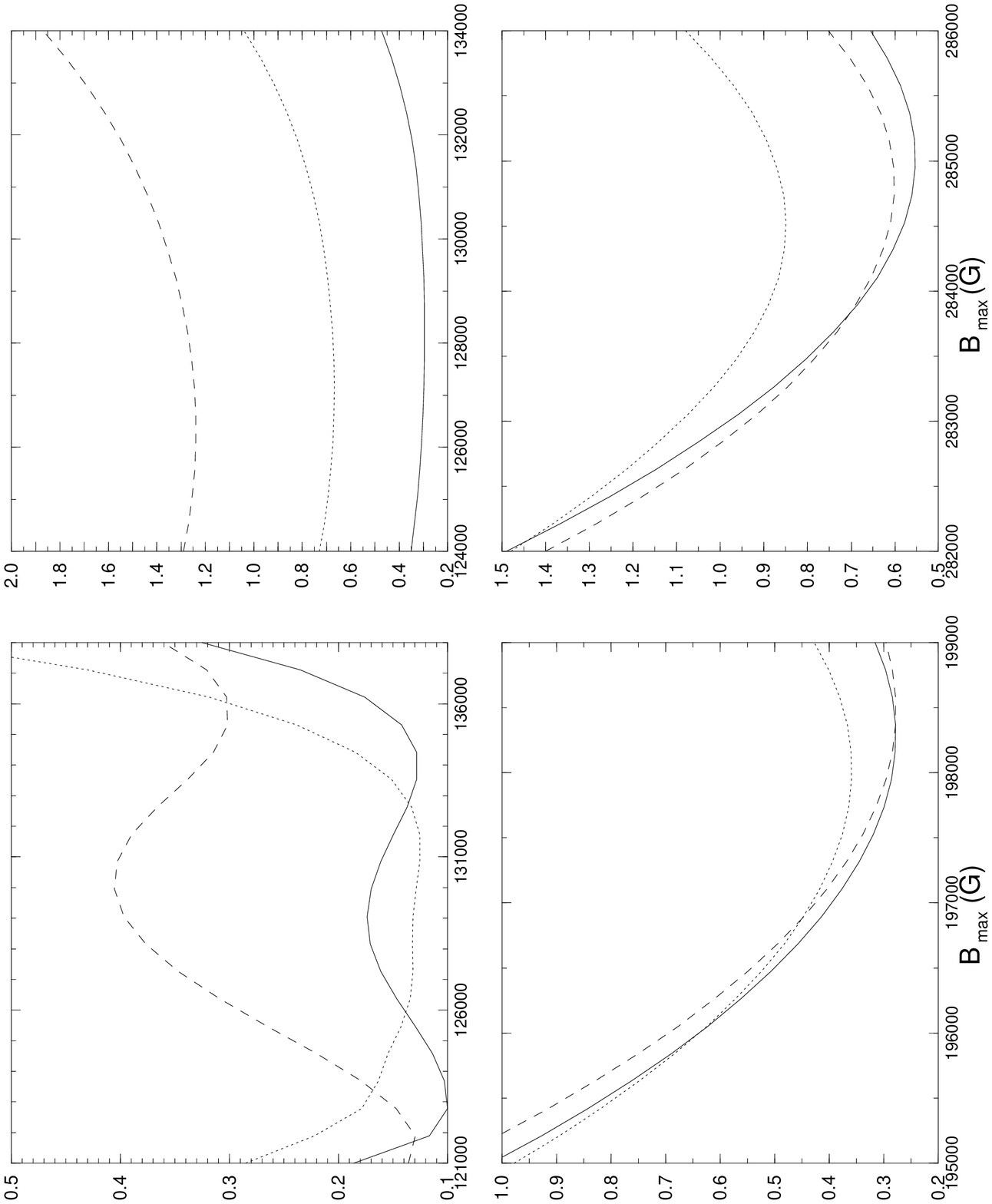,width=15cm}}
\centerline{\mbox{Fig. 6.}}
\end{figure}

\begin{figure}
\mbox{\psfig{figure=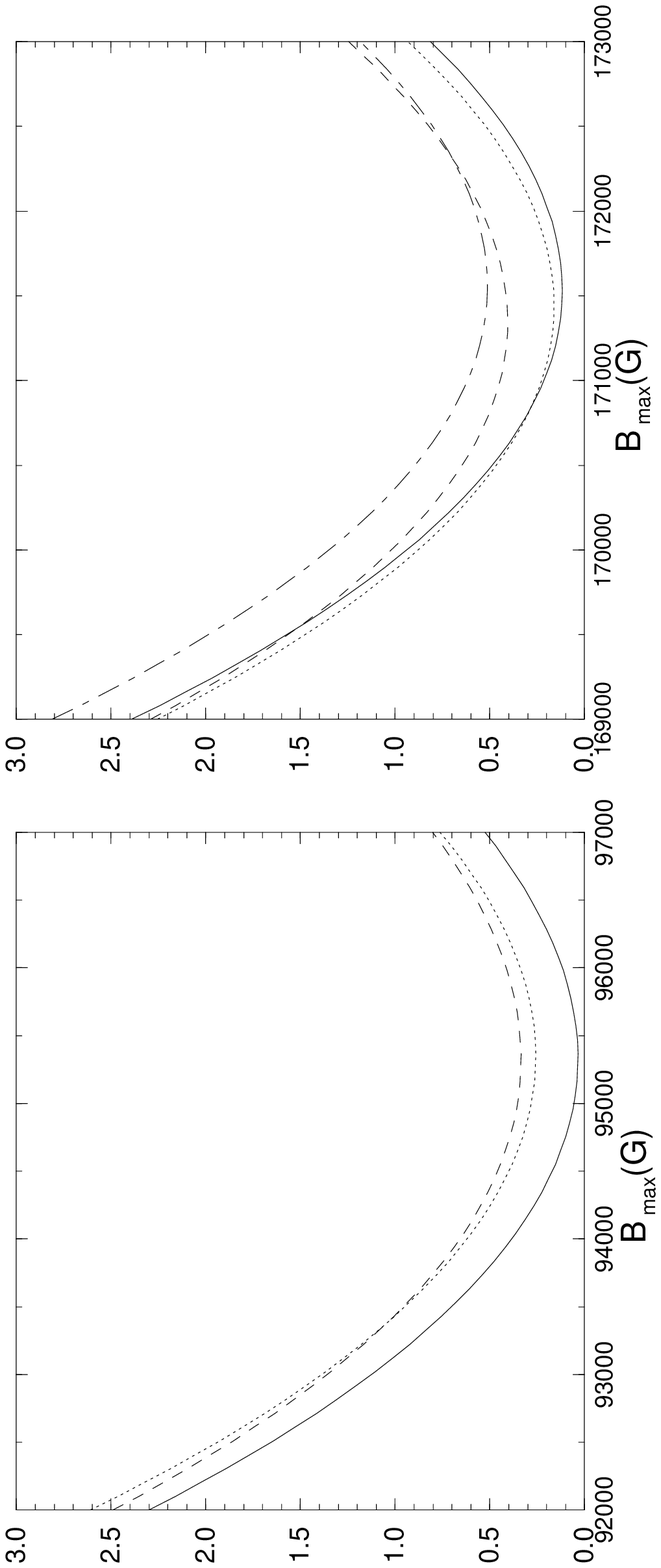,width=15cm}}
\centerline{\mbox{Fig. 7.}}
\end{figure}

\begin{figure}
\mbox{\psfig{figure=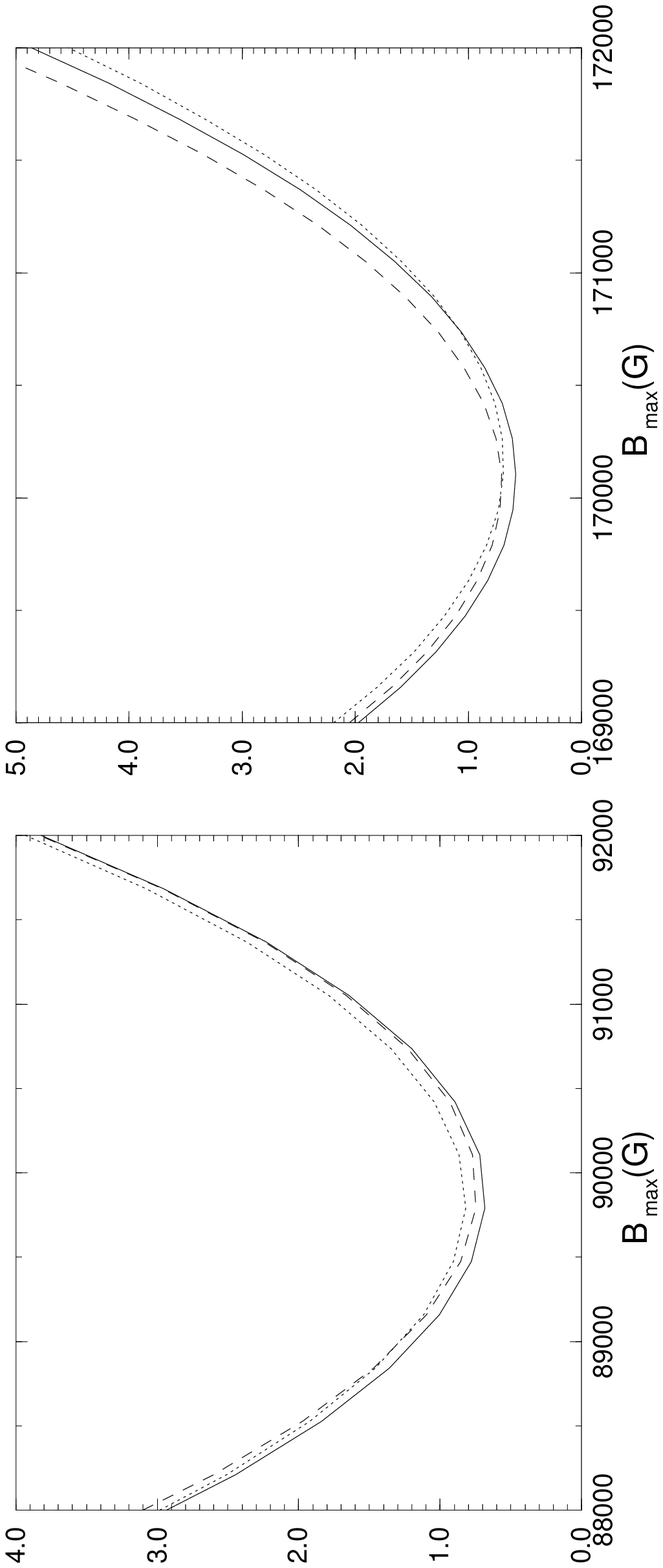,width=15cm}}
\centerline{\mbox{Fig. 8.}}
\end{figure}

\begin{figure}
\mbox{\psfig{figure=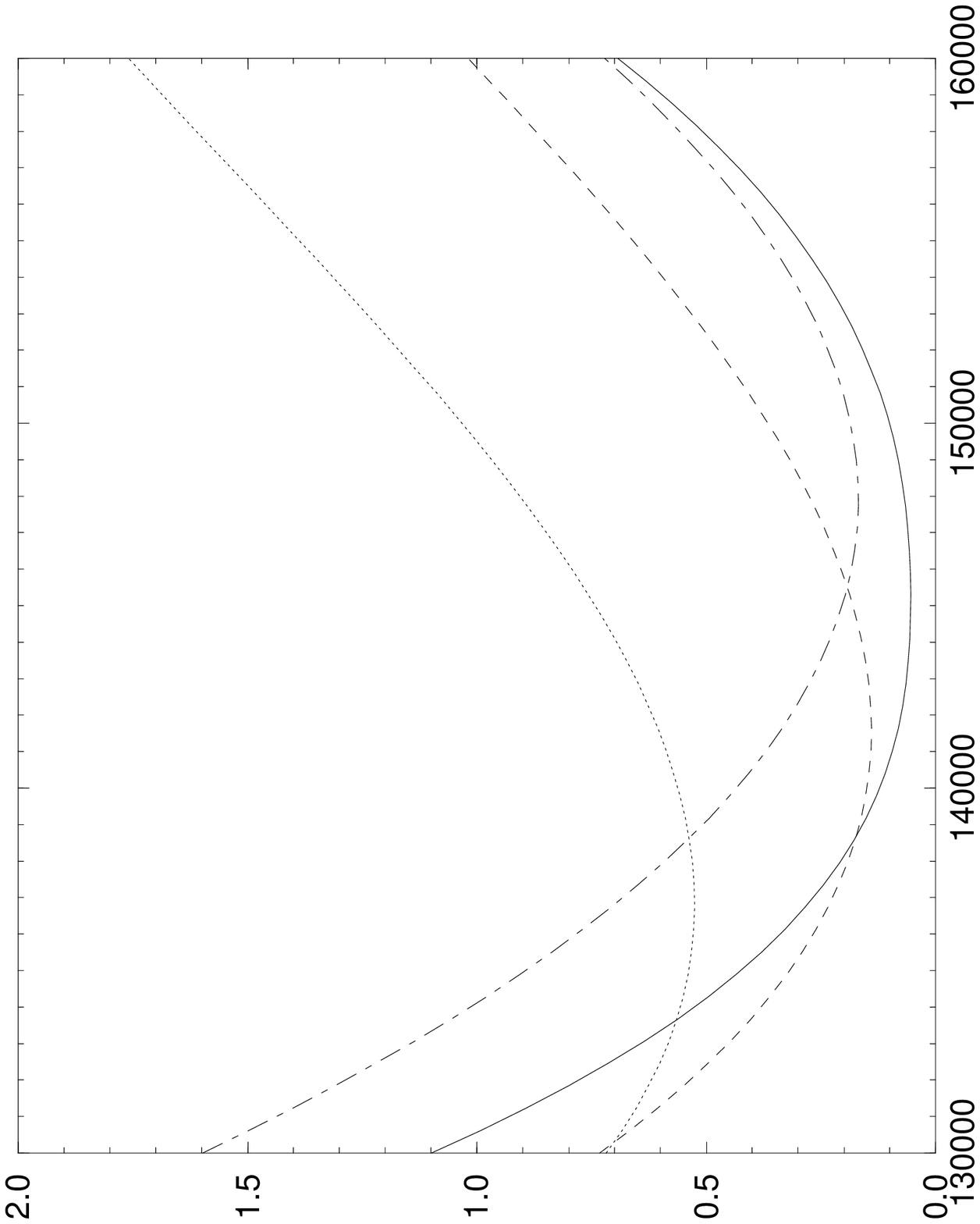,width=15cm}}
\centerline{\mbox{Fig. 9.}}
\end{figure}

\begin{figure}
\mbox{\psfig{figure=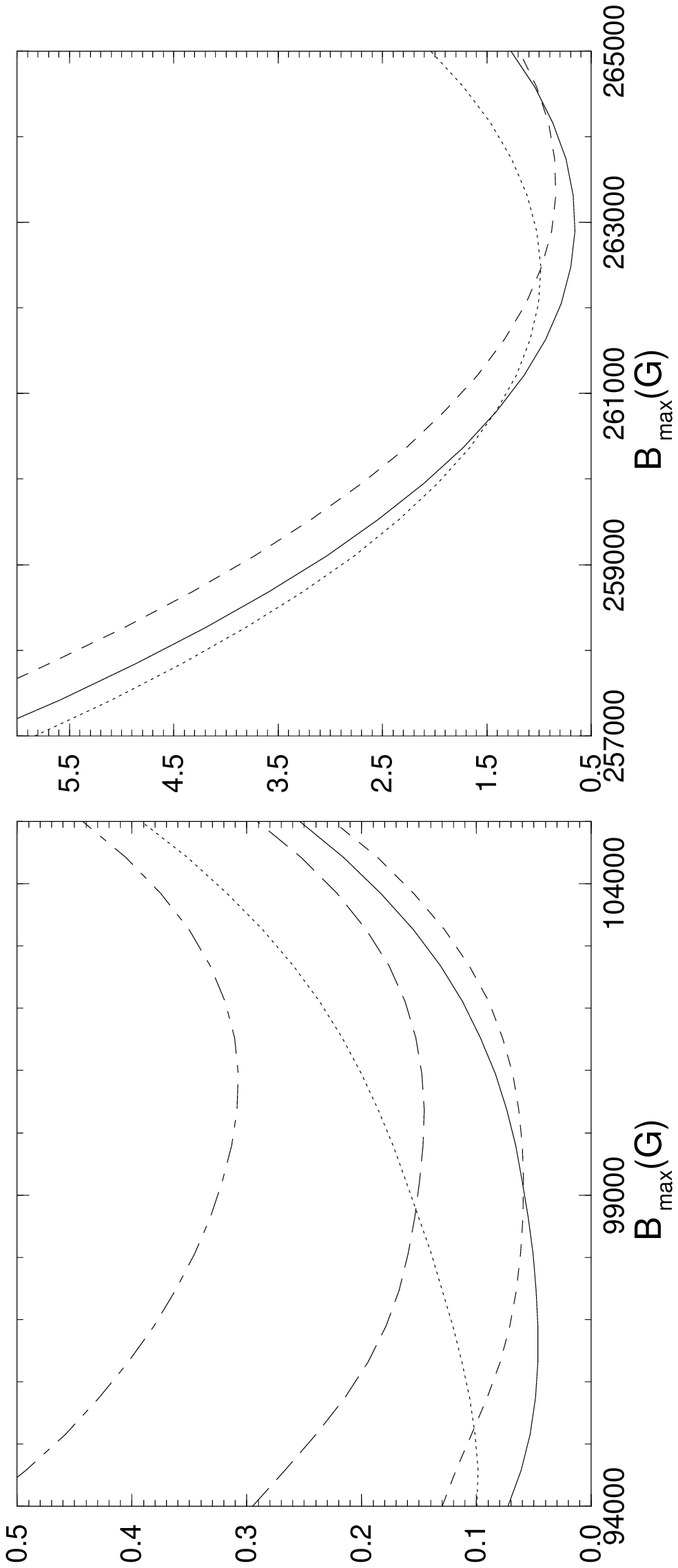,width=15cm}}
\centerline{\mbox{Fig. 10.}}
\end{figure}

\begin{figure}
\mbox{\psfig{figure=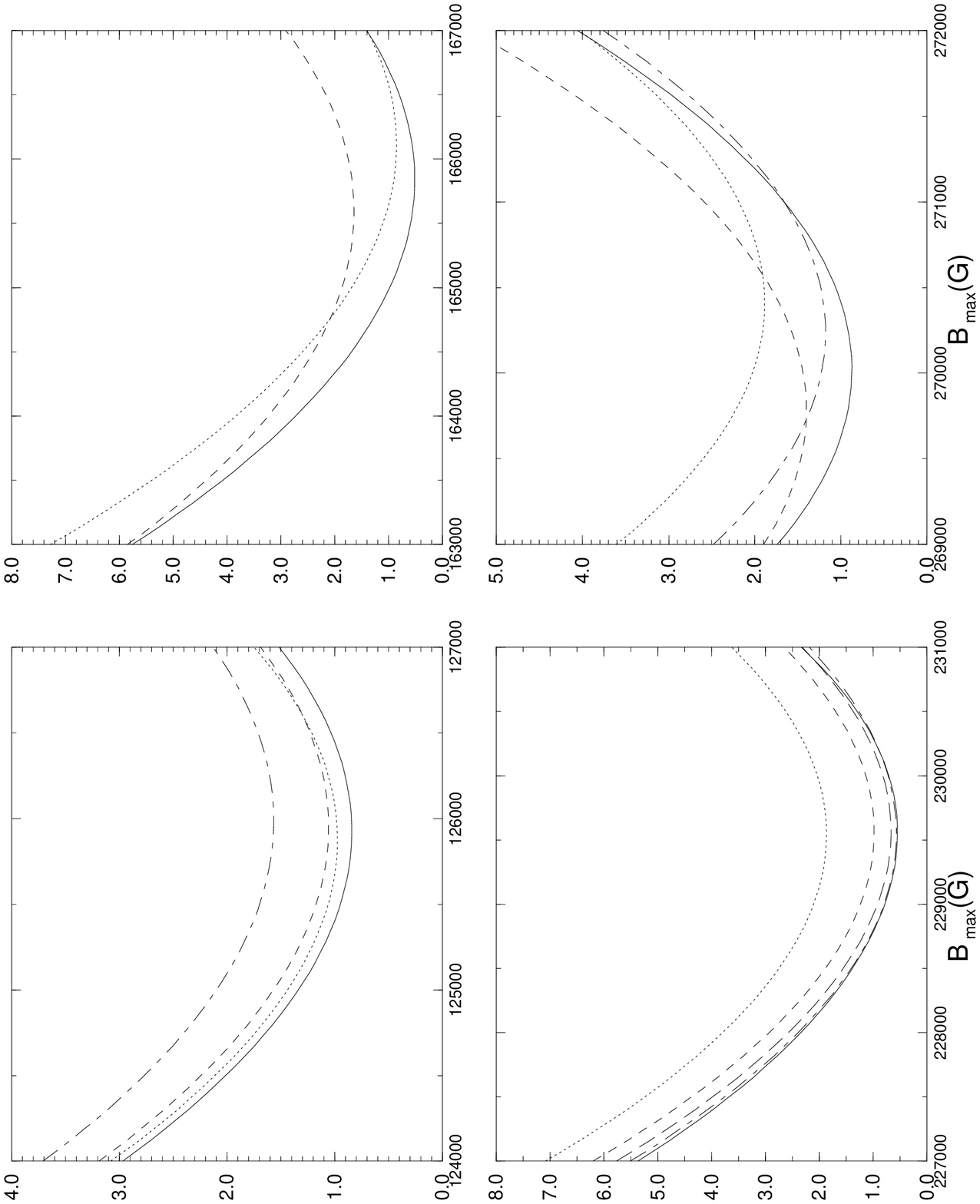,width=15cm}}
\centerline{\mbox{Fig. 11.}}
\end{figure}

\begin{figure}
\mbox{\psfig{figure=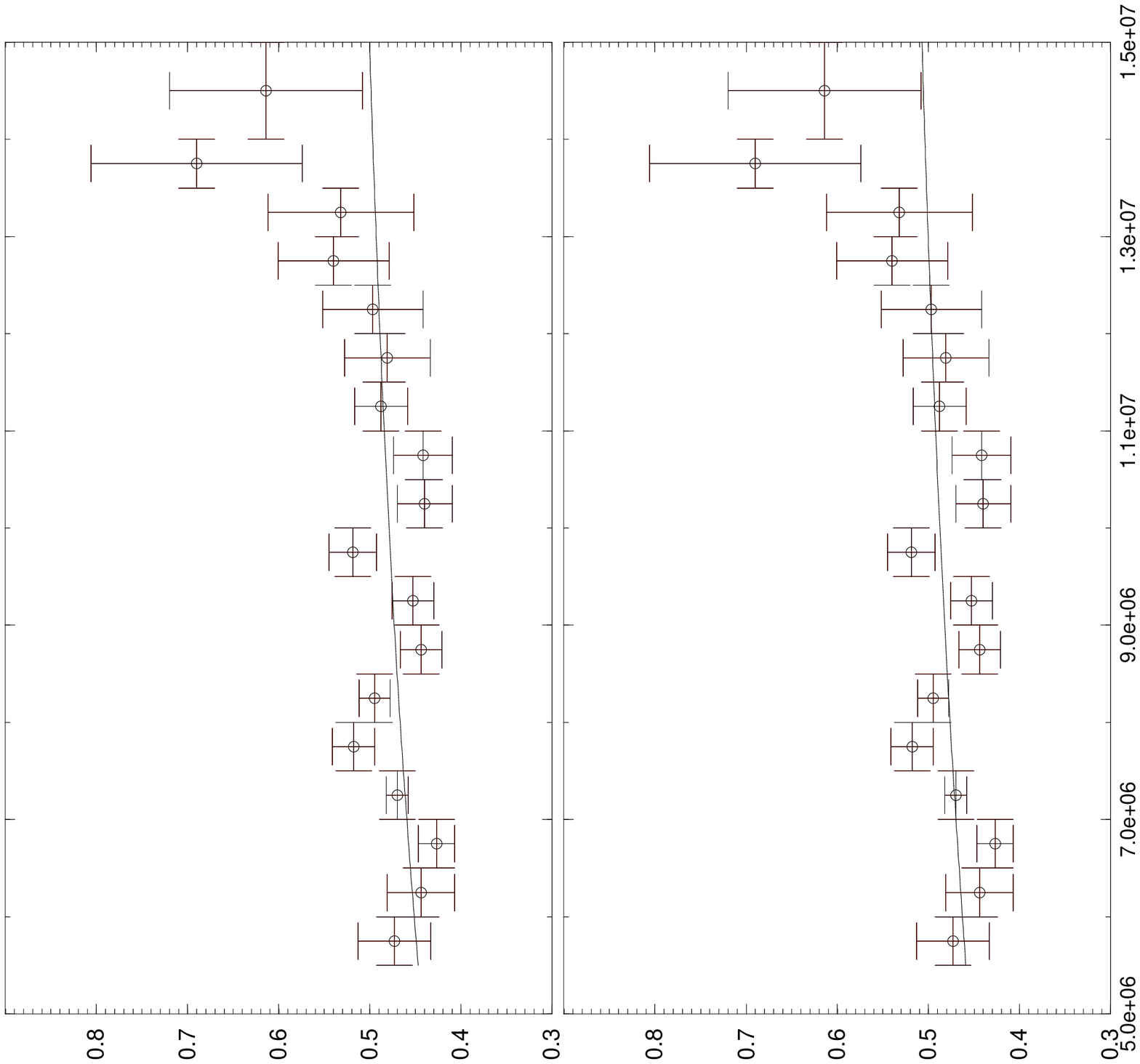,width=15cm}}
\centerline{\mbox{Fig. 12.}}
\end{figure}


\begin{thebibliography}{pppp}
\bibitem{Davis} R. Davis, D. S. Harmer and K. C. Hoffman, Phys. Rev. Lett. {\bf20}, 1205
(1968).
\bibitem{TCL} S. Turck-Chieze and I. Lopes, Astrophys. J. {\bf408}, 347 (1993).
\bibitem{TCCD} S. Turck-Chieze, S. Cahen, M. Casse and C. Doom, Astrophys. J. {\bf335},
415 (1988).
\bibitem{BP92} J. N. Bahcall and M. H. Pinsonneault, Rev. Mod. Phys. {\bf64} (1992) 885.
\bibitem{P} C. R. Profitt, Astrophys. J. {\bf425}, 849 (1994).
\bibitem{RVCD} O. Richard {\it{et al.}}, Astron. Astrophys. {\bf312} 1000 (1996).
\bibitem{Inn} S. Degl'Innocenti {\it{et al.}}, Astron. Astrophys., Suppl. Ser. {\bf123},
1 (1997).
\bibitem{BP95} J. N. Bahcall and M. H. Pinsonneault, Rev. Mod. Phys. {\bf67}, 781 (1995).
\bibitem{BP98} J. N. Bahcall, S. Basu and M. H. Pinsonneault. Phys. Lett. {\bf433} 1 (1998).
\bibitem{BTC} A. S. Brun, S. Turck-Chieze and P. Morel, astro-ph/9806272, to be published in
Astrophys. J. {\bf{506}} (1998).      
\bibitem{Hom} B. T. Cleveland {\it{et al.}}, Astrophys. J. {\bf496}, 505 
(1998); B. T. Cleveland {\it{et al.}}, Nucl. Phys. {\bf B} (Proc. Suppl.) 
{\bf38}, 47 (1995); R. Davis, Prog. Part. Nucl. Phys. {\bf32}, 13 (1994). 
\bibitem{SAGE} SAGE Collaboration, V. Gavrin {\it{et al.}}, in 
{\it{Neutrino 98}}, Proceedings of the XVIII International Conference on
Neutrino Physics and Astrophysics, Takayama, Japan, 4-9 June 1998, edited
by Y. Suzuki and Y. Totsuka. To be published in Nucl. Phys. {\bf B} (Proc.
Suppl.). 
\bibitem{Gallex} Gallex Collaboration, P. Anselmann {\it{et al.}}, Phys. Lett. 
{\bf{B 342}}, 440 (1995); W. Hampel {\it{et al.}}, Phys. Lett. {\bf{B
388}}, 364 (1996). 
\bibitem{SuperK} SuperKamiokande Collaboration, Y. Suzuki in {\it{Neutrino
98}} \cite{SAGE}.  
\bibitem{MSW} L. Wolfenstein, Phys. Rev. {\bf{D 17}}, 2369 (1978); {\bf{20}}, 2634 (1979);
S. P. Mikheyev and A. Smirnov, Sov. J. Nucl. Phys. {\bf{42}}, 913 (1985).
\bibitem{VO} B. Pontecorvo, Zh. Exp. Theor. Fiz. {\bf 53} (1967) 1717; 
V. Gribov and B. Pontecorvo, Phys. Lett. {\bf 28B} (1969) 493; 
V. Barger, R. J. N. Phillips and K. Whisnant, Phys. Rev. {\bf D24} (1981) 538; 
S. L. Glashow and L. M. Krauss, Phys. Lett. {\bf B190} (1987) 199; 
P. I. Krastev and S. T. Petcov, Phys. Rev. {\bf D53} (1996) 1665; 
E. Calabresu et al., Astropart. Phys.  {\bf 4} (1995) 159; 
Z. G. Berezhiani and A. Rossi,  Phys. Rev. {\bf D51} (1995) 5229;
Phys. Lett. {\bf B367} (1996) 219.
\bibitem{hata} N. Hata, P. Langacker, Phys. Rev. {\bf D56} (1997) 6107.
\bibitem{BKS} J. N. Bahcall, P. I. Krastev and A. Yu. Smirnov, Phys. Rev. 
{\bf{D 58}} 096016 (1998).  
\bibitem{BKS2} J. N. Bahcall, P. I. Krastev and A. Yu. Smirnov,
hep-ph/9905220.  
\bibitem{GGHPV} M.C. Gonzalez-Garcia, P.C. de Holanda, C. Pena-Garay and 
J.W.F. Valle, FTUV-99-41, hep-ph/9906469. 
\bibitem{MS}S. P. Mikheyev and A. Yu. Smirnov, Phys. Lett. {\bf B429} (1998) 343; 
\bibitem{BFL} 
V. Berezinsky, G. Fiorentini and M. Lissia, hep-ph/9811352, hep-ph/9904225; 
V. Barger and K. Whisnant, Phys. Lett. {\bf B456} (1999) 54;  
Phys. Rev. {\bf D59}, 093007, 1999. 
\bibitem{Sm} A. Yu. Smirnov, Invited talk at 18th International Conference on
Neutrino Physics and Astrophysics (NEUTRINO 98), Takayama, Japan, 4-9 Jun
1998, hep-ph/9809481. 
\bibitem{ber} V. Berezinsky Talk given at 19th Texas Symposium on Relativistic 
Astrophysics: Texas in Paris, Paris, France, 14-18 Dec. 1998, 
hep-ph/9904259 
\bibitem{LMA} C. S. Lim and W. J. Marciano, Phys. Rev. {\bf{D37}}, 1368 (1988); 
E. Kh. Akhmedov, Sov. J. Nucl. Phys. {\bf{48}}, 382 (1988); 
E. Kh. Akhmedov, Phys. Lett. {\bf{B 213}}, 64 (1988). 
\bibitem{SV}  J. Schechter and J. W. F. Valle, Phys. Rev. {\bf D24}(1981) 1883, 
Erratum-{\it ibid.} {\bf D25} (1982) 283. 
\bibitem{Tom} J. Pulido, Phys. Rev. {\bf{D57}}, 7108 (1998).
\bibitem{GN} M. Guzzo and H. Nunokawa, preprint hep-ph/9810408, Astrop. Phys. 
{\bf 12}, 87 (1999). 
\bibitem{BFZ} 
K.S. Babu and R.N. Mohapatra, Phys. Rev. Lett. {\bf63} (1989) 228; Phys. 
Rev. {\bf D42} (1990) 3778; 
S. M. Barr, E. M. Freire, A. Zee, Phys. Rev. Lett. {\bf{65}}, 2626 (1990).
\bibitem{Parker} E. N. Parker in {\em ``The Structure of the Sun''}, Proc. of the VI 
Canary Islands School, Ed. Roca Cortes and F. Sanchez, Cambridge University Press 
1996 p. 299.
\bibitem{magmo} S. I. Blinnikov, Institute for Theoretical and Experimental
Physics Report No. ITEP-88-19 (1988), unpublished; S. I. Blinnikov,
V.S. Imshennik, D.K. Nadyozhin, Sov. Sci. Rev. {\bf E} Astrophys. Space Sci. {\bf
6},
185 (1987); G.G. Raffelt, Phys. Rev. Lett. {\bf 64} (1990) 2856; Astrophys. J.
{\bf 365} (1990) 559; 
V. Castellani and S. Degl'Innocenti, Astrophys. J. {\bf{402}}, 574 (1993). 
\bibitem{ALP1} E. Kh. Akhmedov, A. Lanza and S. T. Petcov, Phys. Lett. {\bf{B 303}},
85 (1993).
\bibitem{AB} E. Kh. Akhmedov and O.V. Bychuk, Sov. Phys. JETP {\bf} 68 (1989) 250.
\bibitem{ALP2} E. Kh. Akhmedov, A. Lanza and S. T. Petcov, Phys. Lett. {\bf{B 348}},
124 (1995).
\bibitem{PhysRep} See e. g. J. Pulido, Phys. Rep. {\bf{211}}, 167 (1992).
\bibitem{Heid} E. Kh. Akhmedov, Talk given at 4th International Solar Neutrino
Conference, Heidelberg, Germany, 8-11 Apr 1997, hep-ph/9705451.  
\bibitem{hom} J. N. Bahcall's homepage, http://www.sns.ias.edu/~jnb/.
\bibitem{FL} G. L. Fogli and E. Lisi, Astrop. Phys. {\bf{3}}, 185 (1995).
\bibitem{SK} Super-Kamiokande Collaboration, Y. Fukuda {\it et al.}, 
Phys. Rev. Lett. {\bf 81} (1998) 1158; Erratum-{\it ibid.} {\bf 81} (1998) 4279. 
\bibitem{CapeTown} SuperKamiokande Collaboration, Y. Suzuki, Talk at XVII Int. 
Workshop on Weak Interactions and Neutrinos, Cape Town, South Africa, Jan 24-30,
1999.
\bibitem{Ringberg} SuperKamiokande Collaboration, T. Kajita, Talk at 
{\it{Beyond the Desert'99}}, Ringberg Castle, Germany, June 6-12, 1999.
\bibitem{free} R. Escribano, J. M. Frere, A. Gevaert and D. Monderen, Phys. 
Lett. {\bf B444} (1998) 397; J. N. Bahcall and P. I. Krastev, Phys. Lett. 
{\bf B436} (1998) 243.


\end{thebibliography}
\end{document}